\documentclass[conference]{IEEEtran}
\usepackage[utf8]{inputenc}
\usepackage{multicol,url,color}
\usepackage{multirow}
\usepackage{amsmath,amssymb,amsfonts}
\usepackage{algorithm}
\usepackage[noend]{algpseudocode}
\usepackage{graphicx}
\usepackage[multiple]{footmisc}
\usepackage{caption}
\usepackage{bbding}
\usepackage{comment}
\usepackage{textcomp}
\usepackage{pifont}
\makeatother
\usepackage{float}
\usepackage{fancyhdr}
\usepackage{mathtools} 
\usepackage{graphics}
\usepackage{placeins}
\usepackage{array}
\usepackage{enumitem}
 \usepackage{color}
 \usepackage[bookmarks=false]{hyperref}
 \makeatletter

\def\ps@IEEEtitlepagestyle{%
  \def\@oddfoot{\mycopyrightnotice}%
  \def\@evenfoot{}%
}
\def\mycopyrightnotice{%
  {\footnotesize 
  }
  \gdef\mycopyrightnotice{}
}

\@ifundefined{showcaptionsetup}{}{
 \PassOptionsToPackage{caption=false}{subfig}}
\usepackage{subfig}
\makeatother

\usepackage{eso-pic}

\newcommand{\RevisedText}[1]{{\color{black}{#1}}}

\usepackage{tikz}
\newcommand{\toolname}{Force2Vec}

\newcommand{\vect}[1]{\mathbf{#1}}
\newcommand{\Wskip}[1]{ }

\title{Force2Vec: Parallel force-directed graph embedding}

\date{}

\author{\IEEEauthorblockN{Md. Khaledur Rahman}
\IEEEauthorblockA{Indiana University Bloomington\\
morahma@iu.edu}
\and
\IEEEauthorblockN{Majedul Haque Sujon}
\IEEEauthorblockA{Indiana University Bloomington\\
msujon@iu.edu}
\and
\IEEEauthorblockN{Ariful Azad}
\IEEEauthorblockA{Indiana University Bloomington\\
azad@iu.edu}}

\begin{document}
\thispagestyle{plain}
\pagestyle{plain}
\setcounter{page}{1}

\maketitle

\begin{abstract}
     A graph embedding algorithm embeds a graph into a low-dimensional space such that the embedding preserves the inherent properties of the graph.
     While graph embedding is fundamentally related to graph visualization, prior work did not exploit this connection explicitly. 
     We develop Force2Vec that uses force-directed graph layout models in a graph embedding setting with an aim to excel in both machine learning (ML) and visualization tasks.
     We make Force2Vec highly parallel by mapping its core computations to linear algebra and utilizing multiple levels of parallelism available in modern processors.
     The resultant algorithm is an order of magnitude faster than existing methods (43$\times$ faster than DeepWalk, on average) and can generate embeddings from graphs with billions of edges in a few hours.
     In comparison to existing methods, Force2Vec is better in graph visualization and performs comparably or better in ML tasks such as link prediction, node classification, and clustering. \RevisedText{Source code is available at \url{https://github.com/HipGraph/Force2Vec}}.

\end{abstract}


\section{Introduction}
Graphs are powerful models for relational data arisen in social, information, chemical, and biological domains.
The sparsity, irregularity and especially, high dimensionality of real-world graphs made machine learning (ML) on graphs much harder than image and text data.
According to the definition of Erd\"{o}s et al. a graph with $n$ vertices represents an $n$-dimensional feature space~\cite{erdos1965dimension}. 
To learn from such high-dimensional data ($n$ can be billions or more), ML techniques would require enormous amounts of training data to build faithful models. 
To tackle this problem, graphs are often embedded into $d$-dimensional vector space, where $2 \leq d\ll n$.
Thus, graph embedding enables standard ML techniques for classification and regression easily applicable on graphs.

ML methods on graphs can be broadly divided into {\em transductive methods} where all test-time queries are restricted to the set of nodes given in the training data, and {\em inductive methods}  where test-time queries may be defined over  new graphs not available during training.
Inductive algorithms such as GCN~\cite{kipf2016semi} and
GraphSAGE~\cite{hamilton2017inductive}
use neural networks to generalize for unseen data and are more suitable for dynamic graphs. 
By contrast, transductive algorithms such as DeepWalk~\cite{perozzi2014deepwalk}, node2vec~\cite{grover2016node2vec}, Verse~\cite{tsitsulin2018verse}, and HARP~\cite{chen2018harp} directly learn the embedding of graphs in an unsupervised manner. 
Transductive algorithms are often followed by another ML algorithm and are better applicable for static graphs.
This paper only considers transductive graph embedding algorithms. 

Even though graph embedding received more attention in ML research in  recent years, this problem  is well studied by the graph drawing and visualization community for more than 60 years, such as the force-directed algorithm discussed by Tutte in 1963~\cite{tutte1963draw}.
In graph drawing, vertices are embedded in 2D or 3D space so that an ``aesthetically pleasing" layout of the graph can be visualized on a 2D or 3D screen. 
Hence, graph drawing and node embedding both solve the same underlying problem: find a function that maps every vertex  $v$ in a graph to a $d$-dimensional vector $\vect{z}_v{\in}\mathbb{R}^d$.
The primary difference is the value of $d$: graphs drawing uses $d$ = 2 or 3, whereas ML tasks use $d\approx 100$. 
Despite this clear connection, recent graph embedding studies are primarily based on techniques from language models such as word2vec~\cite{mikolov2013distributed}. 
This paper establishes a connection between graph embedding and force-directed graph drawing, thereby bringing ML and visualization lines of research under the same umbrella.

Force-directed layout algorithms consider spring-like attractive forces (e.g., based on Hooke's law) among adjacent vertices and repulsive forces (e.g., based on Coulomb's law between electrically charged particles) among non-adjacent vertices.
By comparison, random-walk-based algorithms such as node2vec perform random walks from a vertex $v$, and vertices reached in these walks form $v$'s context.
Then, these methods form two subsets of vertices based on vertices inside and outside of $v$'s context and maximize a likelihood function. 
Here, we model the likelihood computation within $v$'s context by attractive forces and outside of $v$'s context by repulsive forces.
We develop a novel similarity-based optimization function that combines attractive and repulsive forces to learn a lower dimensional representation.
This expressive force-directed framework named as {\em Force2Vec} opens up the door for all popular graph drawing models applicable to graph embedding. 
We experimentally demonstrate that Force2Vec performs comparably or better for various ML tasks such as node classification and link prediction, and performs better in visualizing graphs. 

Graph embedding algorithms are computationally expensive. 
Most existing algorithms including Force2Vec minimize a loss function by using an optimization method such as Stochastic Gradient Descent (SGD).
Finding embedding of a graph with millions of vertices and hundreds of millions of edges may need hours or even days. For example, parallel DeepWalk needs about a day to find embedding of the Orkut graph (3M vertices and 117M edges) using a 48-core Intel Skylake processor (see Table \ref{tab:runtimes}).
This is a severe impediment in analyzing large-scale social and biological networks.
In this paper, we develop a parallel Force2Vec algorithm that runs an order of magnitude faster than existing methods. 
This speedup comes from two sources: (a) the mathematical formulation of Force2Vec maps the gradient computation to linear algebra operations that are computationally similar to sparse matrix-dense matrix multiplication and (b) we design efficient algorithms for the low-level kernels by employing multiple levels of parallelism and regular data accesses.
Hence, Force2Vec allows us to generate high-quality embedding of large-scale graphs quickly.
The main contributions of this paper are the following:
\begin{itemize}
    \item We merge visualization and ML approaches for graph embedding in the Force2Vec algorithm. Force2Vec achieves state-of-the-art results for  various ML tasks such as node classification and link prediction, and generates high-quality visualization. 
    
    \item Force2Vec provides an expressive framework for graph embedding with mathematical formulation based on linear algebra. Other graph embedding and visualization algorithms can be implemented in this framework.
    
    \item We present a highly-parallel algorithm that uses multicore processors and memory efficiently. Force2Vec is $22\times$ to $56\times$ faster than DeepWalk and $10.5\times$ to $45.4\times$ faster than Verse for various large graphs.
    
    \item Force2Vec can generate embedding of a graph with billions of edges. Most existing algorithms fail to generate embedding for such large graphs.  
\end{itemize}



\section{Background and Related Work}
\subsection{The general structure of the node embedding problem}
Let $G(V,E)$ denote a graph with a set of $n$ vertices $V$ and a set of $m$ edges $E$.
$\vect{A} \in \mathbb{R}^{n\times n}$ denotes the sparse adjacency matrix of the graph where $\vect{A}_{ij} = 1$ if $\{v_i,v_j\}{\in}E$, otherwise $\vect{A}_{ij} = 0$.
The node embedding problem aims to learn a $d$-dimensional embedding matrix $\vect{Z}\in  \mathbb{R}^{n\times d}$ such that every vertex $v$ can be embedded into $d$-dimensional vector space using a function $f$: $v{\rightarrow}\vect{z}_v\in\mathbb{R}^d$, where $\vect{z}_v $ is the embedding of $v$ in $\mathbb{R}^d$ and $d \ll n$.
The embedding function $f$ is also called an {\em encoder} as it encodes a vertex into a vector in $\mathbb{R}^d$.

Let $\delta: \mathbb{R}^d\times \mathbb{R}^d \rightarrow \mathbb{R}$ be a function that computes the {\em similarity} between embedding vectors of two vertices.
In this paper, we consider $\delta$ to be a probability distribution function with $\delta( \vect{z}_u,  \vect{z}_v )$ producing a value between 0 and 1. 
Let $\delta_G(u, v)$ be some proximity between $u$ and $v$ in the original graph.
For example,  $\delta_G(u, v)$ can simply represent vertex connectivity: $\delta_G(u, v) = 1$, when $(u, v)\in E$.
The goal of a graph embedding algorithm is to find an embedding so that $\delta (\vect{z}_u,  \vect{z}_v) \approx \delta_G(u, v)$.
Thus a good embeddings can be found by defining a loss function ${\mathcal{L}(\delta( \vect{z}_u,  \vect{z}_v),  \delta_G(u, v))}$ to measure the discrepancy between the embedded and true proximity values and then minimizing this loss function by using SGD. 
This general structure of embedding algorithms falls into the {\em encoder-decoder} paradigm that is very expressive to capture most unsupervised graph embedding techniques. 
Note that unlike semi-supervised approaches such as GNNs, unsupervised embedding methods need another algorithm such as logistic regression to classify nodes or to predict links.



\subsection{Previous work}

\begin{table*}[!h]
    \centering
    \caption{A summary of state-of-the-art unsupervised models including time and memory complexity, dependency on any specific model in corresponding programming language, and respective parallelization technique. $n$ - the number of vertices, $d$ - embedding dimension, $w$ - walk length, $s$ - the number of negative samples, $l$ - the number of layers in multi-layer graph.}
    \label{tab:minisurvey}
    \begin{tabular}{|c|p{1.6cm}|p{2.0cm}|p{1.3cm}|p{0.9cm}|p{0.95cm}|p{1.0cm}|p{5cm}|}
    \hline
     \textbf{Method}    &  \textbf{Key Focus}  &   \textbf{Similarity}     &  \textbf{Time}   &   \textbf{Mem.}     &   \textbf{Depend.} & \textbf{Language} & \textbf{Parallelization} \\ \hline
     GraRep~\cite{cao2015grarep}   &  Factorization  & $\vect{z}_u^T\vect{z}_v$     & $O(n^3)$  & $O(n^2)$  &  SVD   & MATLAB & \ding{53} \\ \hline
     HOPE~\cite{ou2016asymmetric}   & Factorization   & $\vect{z}_u^T\vect{z}_v$      &  $O(d^2n^2)$  & $O(n^2)$  &  GSVD & MATLAB & \ding{53} \\ \hline
     DeepWalk~\cite{perozzi2014deepwalk}   &  Random Walk  & $\frac{e^{\vect{z}_u^T\vect{z}_v}}{\sum_{k\in V}e^{\vect{z}_u^T\vect{z}_k}}$     & $O(dn\log n)$  & $O(wn^2)$  &    word2vec  & Python  & path sampling by concurrent.futures module and model optimization by word2vec \\ \hline
     LINE~\cite{tang2015line}   &  1$^{st}$/2$^{nd}$ ord. Proximity  & $\frac{1}{1+e^{-\vect{z}_u^T\vect{z}_v}}$    & $O(dsn)$  &  $O(n^2)$  & \ding{53} &  C++ &  inefficiently parallelized using POSIX Threads \\ \hline
     struc2vec~\cite{ribeiro2017struc2vec}   & Random Walk   & $\frac{e^{-f_k(u,v)}}{\sum_{v\in V \atop u \neq v}e^{-f_k(u,v)}}$     &  $O(ln^3)$ &  $O(ln^2)$ &  word2vec    & Python & path sampling by concurrent.futures module and model optimization by word2vec\\ \hline
     Verse~\cite{tsitsulin2018verse}   &  1$^{st}$ Proximity  & $\frac{1}{1+e^{-\vect{z}_u^T\vect{z}_v}}$     & $O(dsn)$  & $O(n^2)$  &  \ding{53} &  C++  &  inefficient parallelization using OpenMP\\ \hline
     HARP~\cite{chen2018harp}   &  Multilevel  & model dependent    & $O(dsn)$  & $O(n^2)$   & word2vec  &  Python &  heavily dependent on underlying model \\ \hline
    \end{tabular}
     \vspace{-0.3cm}
\end{table*}
Graph embedding is a well-studied problem in graph mining and machine learning literature.
We refer readers to a recent survey~\cite{cai2018comprehensive} for a comprehensive view of the field.
Researchers have previously developed various encoding and decoding schemes, graph proximity measures, and loss functions~\cite{perozzi2014deepwalk,grover2016node2vec,tsitsulin2018verse,tang2015line}.
We can broadly categorize them based on the encoding schemes and the learning strategies. We provide a brief summary of some embedding methods in Table \ref{tab:minisurvey}.

Early methods of graph mining perform manual feature extraction from graphs such as degree, clustering co-efficient, etc. which can not fully capture the inherent structure of the graph \cite{henderson2012rolx,akoglu2010oddball}. 
Later, matrix factorization based methods~\cite{ahmed2013distributed, cao2015grarep} have been introduced to decompose graphs (represented by various matrices) using Singular Value Decomposition (SVD) or Non-negative Matrix Factorization (NMF). Lack of parallelization and high runtime/memory complexity restrict them to be applied on bigger graphs.


A plethora of random-walk based methods have been introduced for graph representation learning. 
DeepWalk\cite{perozzi2014deepwalk} performs random-walk on the graph to sample a set of paths for each vertex. 
Then, the problem is formulated using the \emph{word2vec} model~\cite{mikolov2013distributed} to generate embedding of nodes.
\emph{LINE} captures information from the graph for first order and second order proximities i.e, 1-hop and 2-hop neighbors \cite{tang2015line}. The
node2vec method samples walks based on Depth First Search (DFS) and Breadth First Search (BFS) traversal \cite{grover2016node2vec}. Tsitsulin et al. introduce a versatile graph embedding method, called Verse, that uses various approaches to instantiate the embedding \cite{tsitsulin2018verse}. Verse shows that the stationary distribution of random walk eventually converges to personalized pagerank \cite{page1999pagerank}. 
HARP \cite{chen2018harp} developed multi-level algorithms for three state-of-the-art methods~\cite{walshaw2006multilevel}. 
Even though HARP made some embedding methods faster, it consumes a significant amount of memory for storing several intermediate graphs.


Generally, \RevisedText{a} general purpose graph embedding method should run fast, consume less memory and generate high quality embeddings that perform well in different prediction tasks. 
To accomplish these goals, we introduce an unsupervised parallel algorithm called \toolname{}.

\section{The Force2Vec Algorithm}
A good embedding of a graph preserves structural information from the original graph in the embedding space. 
Thus, the neighboring vertices usually have ``similar" embeddings than non-neighboring vertices.
Let $N(u)$ denote the ``neighbors" of $u$ in a particular context. 
Here, $N(u)$ can simply denote vertices connected with $u$ by edges or vertices reached by random walks from $u$.
Using the similarity distribution function $\delta$, we can define the loss function for vertex $u$ as the negative log likelihood with respect to all other vertices in the graph:
\begin{equation}
\label{eqn:logloss_nosample}
    \mathcal{L}(u) = -\sum_{\mathclap{v\in N(u)}}\log \delta(\vect{z}_u, \vect{z}_v) - \sum_{\mathclap{w\notin N(u)}}\log (1-\delta(\vect{z}_u, \vect{z}_{w})).
\end{equation}
We can then optimize Equation~\ref{eqn:logloss_nosample} by minimizing $\sum_{u\in V}\mathcal{L}(u)$ using the minibatch Stochastic Gradient Descent (SGD) algorithm. 
Similar to prior work~\cite{grover2016node2vec, tsitsulin2018verse}, we use negative samples to reduce the cost of the second term in Equation~\ref{eqn:logloss_nosample}.
Let $S(u)$ denote a subset of vertices considered as negative samples for $u$. Then we use the following equation throughout the paper:
\begin{equation}
\label{eqn:logloss}
    \mathcal{L}(u) = -\sum_{\mathclap{v\in N(u)}}\log \delta(\vect{z}_u, \vect{z}_v) - \sum_{\mathclap{w\in S(u)}}\log (1-\delta(\vect{z}_u, \vect{z}_{w})).
\end{equation}

\subsection{Force model of the loss function}
In this paper, we recast the loss function in terms of a spring-electrical model of force-directed graph drawing where two types of forces are calculated based on the connectivity of the vertices, namely, an \emph{attractive} force when two vertices are connected by an edge and a \emph{repulsive} force when there is no connection between a pair of vertices~\cite{fruchterman1991graph,rahman2020batchlayout}. 
With a good choice of the similarity function $\delta$, the first term in Eq.~\ref{eqn:logloss} can be considered as the attractive force between a pair of vertices and the second term can be considered as the repulsive force.
Next we discuss several similarity functions that would generate \RevisedText{embeddings that are effective for visualization and other downstream machine learning tasks}.

{\bf The sigmoid function.} 
Our first similarity function is based on the sigmoid function applied on the dot product of two embedding vectors~\cite{perozzi2014deepwalk,grover2016node2vec,tsitsulin2018verse}.
Using $\delta(\vect{z}_u, \vect{z}_v) = \sigma(\vect{z}_u\vect{z}_v) = \frac{1}{1+e^{-\vect{z}_u\vect{z}_v}}$ in Eq.~\ref{eqn:logloss}, we get the following loss function:
\begin{equation}
\label{eqn:sigmoidloglikelihood}
\vspace{-0.2cm}
    \mathcal{L}(u) = \sum_{\mathclap{v\in N(u)}}\log(1+e^{-\vect{z}_u\vect{z}_v}) - \sum_{\mathclap{w\in S(u)}}\log \frac{e^{-\vect{z}_u\vect{z}_{w}}}{1+e^{-\vect{z}_u\vect{z}_{w}}}
\end{equation}

In Eq. \ref{eqn:sigmoidloglikelihood}, the first part represents an attractive force which is denoted by $f_a(u) = \sum_{v\in N(u)}\log(1+e^{-\vect{z}_u\vect{z}_v})$ and the second part represents a repulsive force which is denoted by $f_r(u) =-\sum_{w\in S(u)}\log \frac{e^{-\vect{z}_u\vect{z}_{w}}}{1+e^{-\vect{z}_u\vect{z}_{w}}}$. 
Hence, the combined force for the vertex $u$ is represented by:
\begin{equation}
\label{eqn:combinedloss}
\vspace{-0.2cm}
    \mathcal{L}(u) = f_a(u) + f_r(u).
\end{equation}
To compute the gradient of Eq.~\ref{eqn:sigmoidloglikelihood}, we find the derivative of $f_a(u)$ and $f_r(u)$ with respect to $\vect{z}_u$ as follows:
\begin{equation}
\label{eqn:faderivativesigmoid}
    \nabla f_a(u) = \frac{\partial f_a(u)}{\partial \vect{z}_u} = -\sum_{\mathclap{v\in N(u)}} (1-\sigma(\vect{z}_u\vect{z}_v)).\vect{z}_v
\end{equation}

\begin{equation}
\label{eqn:frderivativesigmoid}
   \nabla f_r(u) = \frac{\partial f_r(u)}{\partial \vect{z}_u} = \RevisedText{ \sum_{\mathclap{w\in S(u)}} \sigma(\vect{z}_u\vect{z}_w).\vect{z}_w}
\end{equation}
\RevisedText{Let $\nabla f_a(u,v) = (\sigma(\vect{z}_u\vect{z}_v)-1)$ be the attractive gradient betwenn $u$ and $v$, and $\nabla f_r(u,w) = \sigma(\vect{z}_u\vect{z}_w).\vect{z}_w$ be the repulsive gradient between $u$ and $w$.} 
Then,  $\nabla f_a(u) = \sum_{v\in N(u)}{\nabla f_a(u,v)}$ and $\nabla f_r(u) = \sum_{w\in S(u)}{\nabla f_r(u,w)}$.

{\bf Student's $t$-distribution.}
Our second similarity function is $t$-distribution which has been found effective in high-dimensional data visualization~\cite{maaten2008visualizing,tang2016visualizing}. 
Note that $t$-distribution with one degree of freedom approximates a standard Cauchy distribution \cite{luo2011cauchy}.
Putting the value of $\delta(\vect{z}_u, \vect{z}_v) = \frac{1}{1+t^2}$, where $t = \parallel\vect{z}_u - \vect{z}_v\parallel$, in Eq. \ref{eqn:logloss}, we get the following:
\begin{equation}
\label{eqn:tdistloss}
    \mathcal{L}(u) = \sum_{\mathclap{v\in N(u)}}\log(1+t^2) - \sum_{\mathclap{w\in S(u)}}\log \frac{t^2}{1+t^2}
\end{equation}

We can compute the gradient similarly to the sigmoid function:  
\begin{equation}
\label{eqn:fatgrad}
    \nabla f_a(u) = \frac{\partial f_a(u)}{\partial \vect{z}_u} = \frac{\partial f_a(u)}{\partial t} \frac{\partial t}{\partial \vect{z}_u} = \sum_{\mathclap{v\in N(u)}} \frac{2t}{1+t^2}
\end{equation}
\begin{equation}
\label{eqn:frtgrad}
    \nabla f_r(u) = \frac{\partial f_r(u)}{\partial \vect{z}_u} = \frac{\partial f_r(u)}{\partial t} \frac{\partial t}{\partial \vect{z}_u}= \sum_{\mathclap{w\in S(u)}} \frac{-2}{t(1+t^2)}
\end{equation}

{\bf Other force-directed models.}
In addition to Eqs. \ref{eqn:fatgrad} and \ref{eqn:frtgrad}, we have also employed some other popular force-directed models as gradient components, which are summarized in Table \ref{tab:diffmodels}. Most of these models are very popular in the graph drawing community and the authors of ForceAtlas2 provide a precise description~\cite{jacomy2014forceatlas2}. 

\begin{table}[!ht]
\centering
\caption{Different spring-electrical model with corresponding gradients of attractive and repulsive forces.}
\begin{tabular}{|c|c|c|}

\hline
\textbf{\begin{tabular}[c]{@{}c@{}}Spring-Electrical Model\end{tabular}} & \textbf{\begin{tabular}[c]{@{}c@{}}$\nabla f_a(u,v)$\end{tabular}} & \textbf{\begin{tabular}[c]{@{}c@{}}$\nabla f_r(u,w)$\end{tabular}} \\ \hline
\begin{tabular}[c]{@{}c@{}}Fruchterman Reingold\end{tabular}            & -$\parallel \vect{z}_u - \vect{z}_v\parallel^2$                                                                  & $1/\parallel \vect{z}_u - \vect{z}_w\parallel$\\ \hline
LinLog                                                                     & -$\log(1+\parallel \vect{z}_u - \vect{z}_v\parallel)$                                                            & $1/\parallel \vect{z}_u - \vect{z}_w\parallel$                                                                \\ \hline
ForceAtlas                                                                 & -$\parallel \vect{z}_u - \vect{z}_v\parallel$                                                                   & $1/\parallel \vect{z}_u - \vect{z}_w\parallel$                                                                \\ \hline
\end{tabular}
\label{tab:diffmodels}
\vspace{-0.45cm}
\end{table}

\subsection{Optimization}
After we have the gradient, we can optimize the embedding using the SGD algorithm. 
We will update the embedding in each iteration of the SGD for vertex $u$ as follows:
\begin{equation}
\label{eqn:update}
    \vect{z}_u = \vect{z}_u - \eta\frac{\partial\mathcal{L}(u)}{\partial \vect{z}_u}
\end{equation}
In Eq.~\ref{eqn:update}, $\eta$ is the learning rate or step value. Notably, when we use $t$-distribution, we will need to multiply Eqs. \ref{eqn:fatgrad} and \ref{eqn:frtgrad} by a unit vector of the corresponding embedding of associated vertices to get the direction of movement.

{\bf Minibatch SGD.}
A vanilla SGD that processes just one vertex in each update in  Eq.~\ref{eqn:update} does not offer enough parallelism for multiple threads. 
In a minibatch SGD, we compute the gradient for a minibatch of vertices and update the embedding of the minibatch in parallel.
Here, we only implement the synchronous version of minibatch-SGD that processes each vertex in a minibatch independently and thus provides deterministic results, unlike asynchronous parallelization of SGD \cite{recht2011hogwild}.


\begin{algorithm}[!t]
\caption{Negative Sampling based Force2Vec Algorithm}\label{euclid}
\begin{flushleft}
\textbf{Input:} G(V, E) and an initial embedding $\vect{Z}$\\
\textbf{Output:} optimized embedding $\vect{Z}$
\end{flushleft}
\begin{algorithmic}[1]
\State $\eta = 0.02$ \Comment{initial learning rate}
\For{$c \leftarrow$ 0 to $nepoch-1$}
    \State Partition $V$ into $\lceil n/b \rceil$ minibatches such that each minibatch $V_b$ has approximately $b$ vertices in it

    \For{each minibatch of vertices $V_b$}
        \State $N_b \gets $ neighbors of vertices in $V_b$
        \State $S_b \gets $ Random negative samples; \RevisedText{$|S_b|=s$}  
        
        \State $\nabla f$ = \Call{GradMinibatch}{$G, \vect{Z}, V_{b}, N_b, S_b$}

        \For{$u \in V_{b}$ {\bf in parallel}}
            \State $\vect{z}_u = \vect{z}_u - \eta \times \nabla f(u)$
        \EndFor
    \EndFor
\EndFor
\State \Return the final embedding $\vect{Z}$
\end{algorithmic}
\label{algo:nsforce2vecalgo}
\end{algorithm}

\begin{algorithm}[!t]
\caption{Gradient computation for a minibatch based on attractive and repulsive forces}\label{euclid}
\begin{flushleft}
\textbf{Input:} $G$: the input graph, $\vect{Z}$: the current embedding matrix\\
$V_b$: the set of vertices in the current minibatch; $|V_b|=b$\\ 
$N_b$: the set of neighbors for computing attractive forces\\ 
$S_b$: the set of sampled vertices for computing repulsive forces \\
\textbf{Output:} $\nabla f$: a $b\times d$ matrix, each row stores the gradient for a vertex in $V_b$. $\nabla f(u)$ is the gradient for the vertex $u$.  

\end{flushleft}
\begin{algorithmic}[1]
\Procedure{GradMinibatch}{$G, \vect{Z}, V_b, N_b, S_b$}
    \For{$u \in V_b$ {\bf in parallel}}
        
        \State $\nabla f_a(u) \gets 0;  \nabla f_r(u) \gets 0$ 
        \For{$v \in N_b(u)$ } \Comment{based on attractive force}
            \State $\nabla f_a(u) \gets \nabla f_a(u) + \nabla f_a(u,v)$ Eq. \ref{eqn:faderivativesigmoid} or \ref{eqn:fatgrad}
        \EndFor
        \For{ \RevisedText{$w \in S_b$} } \Comment{based on repulsive force}
            \State $\nabla f_r(u) \gets \nabla f_r(u) + \nabla f_r(u,w)$ Eq. \ref{eqn:frderivativesigmoid} or \ref{eqn:frtgrad}
        \EndFor
    \State $\nabla f(u) \gets \nabla f_a(u)  + \nabla f_r(u) $ 
    \EndFor
\State \Return $\nabla f$
\EndProcedure
\end{algorithmic}
\label{algo:forceforbatch}
\end{algorithm}

\subsection{The Force2Vec Algorithm}
Algorithm~\ref{algo:nsforce2vecalgo} provides a generic description of the \toolname{} algorithm based on negative sampling.
This algorithm can be easily adapted based on different force models.
Line 3 creates minibatches of equal sizes (randomly without replacement).
Each iteration (line 4-9) of Algorithm~\ref{algo:nsforce2vecalgo} computes the gradient of a minibatch of $b$ vertices $V_b$ and then updates the embedding of vertices in $V_b$.
For a minibatch $V_b$, we identify a set $N_b$ of neighboring vertices following a given strategy.
For example, $N_b$ can be vertices adjacent to $V_b$ or vertices discovered by random walks from $V_b$.
Hence, the attractive forces are computed between \RevisedText{$V_b$ and $N_b$}.
\RevisedText{When computing repulsive forces, we use a negative sampling approach~\cite{mikolov2013distributed,grover2016node2vec}.
A set of negative samples for a vertex is formed with a subset of non-neighboring vertices that are chosen randomly from a uniform distribution.}
In line 6 of Algorithm~\ref{algo:nsforce2vecalgo}, we choose a set of negative sample $S_b$ for the current minibatch.
\RevisedText{$S_b$ contains $s$ vertices that are used by every vertex in the minibatch to compute repulsive forces.} 
After $N_b$ and $S_b$ are formed, we can compute gradients (line 7) based on our force model as discussed in the next section. 
Lines 8 and 9 update the embedding of vertices in $V_b$ using the computed gradient.

{\bf Gradient computation based on force models.}
Algorithm~\ref{algo:forceforbatch} describes how we compute gradients for each vertex in the current minibatch $V_b$.
Algorithm~\ref{algo:forceforbatch} also takes $N_b$ (neighbors of $V_b$) and $S_b$ (negative samples for $V_b$) as inputs. 
We use $N_b$ to compute attractive forces and $S_b$ to compute repulsive forces, both with respect to $V_b$.
Each iteration of the outer for loop at line 2 computes the gradient $\nabla f(u)$ for a vertex $u\in V_b$. 
Since gradients for vertices in $V_b$ are independently computed, several threads can process vertices in parallel. 
Lines 4 to 5 of Algorithm~\ref{algo:forceforbatch} show the gradient computation based on the attractive force with respect to vertices in  $N_b$.
Specifically, line 5 computes $\nabla f_a(u,v)$ between $u$ and its neighbor $v$ in $N_b$.  
This gradient computation ($\nabla f_a(u,v)$) depends on the force model such as based on Eqs. \ref{eqn:faderivativesigmoid} or \ref{eqn:fatgrad}.
Similarly, lines 6 to 7 of Algorithm~\ref{algo:forceforbatch} show the gradient computation based on the repulsive force with respect to vertices in  $S_b$.
Specifically, line 6 computes $\nabla f_r(u,w)$ between $u$ and its negative sample $w$ in $S_b$.  
Line 8 adds the gradient components from attractive and repulsive forces. 
Finally, the function returns a $b\times d$ matrix storing gradients for each vertex in $V_b$.


{\bf Initialization and hyper parameters.}
Algorithm~\ref{algo:nsforce2vecalgo} starts with a random embedding $\vect{z}_u$ for every vertex $u$, an initial learning rate $\eta$, size of a minibatch $b$ and the number of iterations $nepoch$. 
These parameters can be tuned empirically as shown in the result section.


{\bf Computational complexity.}
Each gradient component $\nabla f_a(u,v)$ and $\nabla f_r(u,w)$ can be computed by operations on the embedding vectors $\vect{z}_u$, $\vect{z}_v$, and $\vect{z}_w$.
Hence, the complexity of computing $\nabla f_a(u,v)$ and $\nabla f_r(u,w)$ is $O(d)$.
If we consider just 1-hop neighbors in $N_b$, then  $\nabla f_a(u,v)$ is needed to be computed for every edge in the graph, giving us $O(md)$ complexity for the attractive force computation.
This complexity can change if multi-hop neighbors or random walks are used.
\RevisedText{Each vertex computes repulsive forces with respect to $s$ negative samples}, 
giving us an overall cost of $O(nsd)$ for the repulsive force computations.
Hence, per-iteration complexity of \toolname{} is $O(md+nsd)$. 
Here, the relative cost of attractive and repulsive forces depends on the neighborhood formation and negative sampling strategies.


\begin{algorithm}[!ht]
\caption{Random Walk Generation}\label{euclid}
\begin{flushleft}
\textbf{Input:} G(V, E), $V_b$: the set of vertices in the current minibatch, and $k$: the length of walks\\
\textbf{Output:} $N_b$: the set of vertices reached by random walks 
\end{flushleft}
\begin{algorithmic}[1]

\Procedure{GenWalk}{$G, V_b, k$}
    \For{$u \in V_{b}$ {\bf in parallel}}
        \State $N_b(u) \gets \phi$, \RevisedText{$w \gets u$}
        
        \While{$|N_b(u)| < $ $k$}
     
            \State $\RevisedText{w_{next}} \leftarrow $ a random neighbor of $\RevisedText{w}$
            \State $N_b(u) \gets N_b(u) \cup \{\RevisedText{w_{next}}\}$
            \State $\RevisedText{w \gets w_{next}}$

        \EndWhile
    \EndFor
\State \Return $N_b$
\EndProcedure
\end{algorithmic}
\label{algo:nsforce2vecalgorw}
\end{algorithm}

{\bf r\toolname{}: Random-walk based \toolname{}.}
Algorithm~\ref{algo:nsforce2vecalgo} provides a generic framework for \toolname{} and can be easily adapted for other force and embedding models. 
For example, we can form neighborhood of a vertex based on random walks as was used in DeepWalk and node2vec.
To capture this model, we need to create $N_b$ by random walks from the current minibatch $V_b$ as shown in Algorithm~\ref{algo:nsforce2vecalgorw}.
For each vertex $u$ (lines 4 to 7), we select $k$ vertices from its subsequent $k$-hop neighbors for calculating attractive forces. 
After $N_b$ is created using Algorithm~\ref{algo:nsforce2vecalgorw}, we can pass it to Algorithm \ref{algo:nsforce2vecalgo} without changing the gradient computations.
We call this approach r\toolname{} that may perform better for heterogeneous networks by extracting information from multi-hop neighbors. 
The complexity of computing the attractive force is $O(nkd)$.
The complexity of computing the repulsive force in r\toolname{} remains the same.

\begin{figure}[!t]
    \centering
    \fbox{\includegraphics[width=0.95\linewidth]{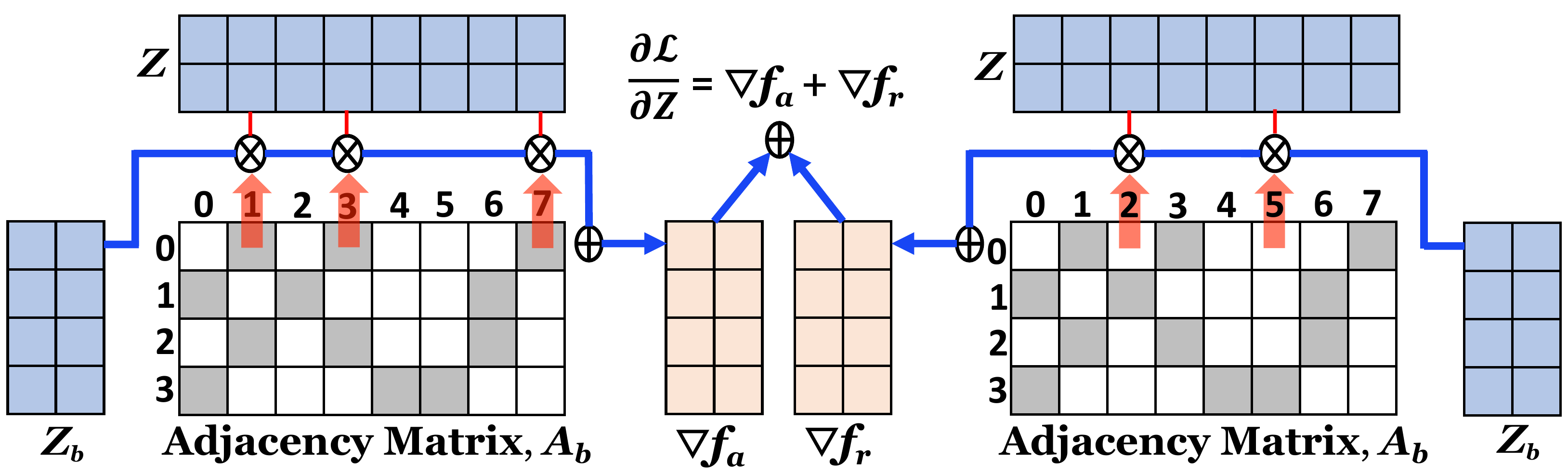}}
    \caption{Gradient computations in linear-algebraic format. The minibatch consists of four vertices. $\nabla f_a$ and $\nabla f_r$ denote gradients with respect to attractive and repulsive forces. Red arrows denote computations guided by the adjacency matrix. }
    \label{fig:gradla}
    \vspace{-0.4cm}
\end{figure}

\subsection{Parallel \toolname{}}
{\bf A linear algebra model.} 
The gradient computations based on attractive and repulsive forces (that is, Algorithm~\ref{algo:forceforbatch}) dominate the runtime of \toolname{}. 
To effectively optimize gradient computations, we model them as a series of linear algebra operations. 
Let $V_b$ be the set of $b$ vertices in the current minibatch and $\vect{Z}_b \in  {\rm \mathbb{R}}^{b\times d}$ be the slice of the embedding matrix corresponding to $V_b$.
Also assume that $\vect{A}_b \in  {\rm \mathbb{R}}^{b\times n}$ denotes the slice of the adjacency matrix storing the edges corresponding to $V_b$.
Figure~\ref{fig:gradla} shows an example of this setting where the minibatch consists of the first four vertices. 

To compute the attractive force based on Eq.~\ref{eqn:faderivativesigmoid}, we need to operate on the embeddings of $V_b$ and the embeddings of their neighbors $N_b$.
For example, in Fig.~\ref{fig:gradla},  vertex $v_0$ in row0 of the adjacency matrix has three non-zero elements at positions 1, 3, and 7, which means $N_b(v_0) = \{v_1,v_3,v_7\}$.
Hence, we access $\vect{z}_1$, $\vect{z}_3$, and $\vect{z}_7$ (shown in red arrows in Fig.~\ref{fig:gradla}), perform 
some computations based on force equations (e.g., Eq.~\ref{eqn:faderivativesigmoid}) 
and then, sum them up to compute the gradient $\nabla f_a(v_0)$ for vertex $v_0$.
This computation follows the pattern of a sparse-dense matrix multiplication (SpMM).
Similarly, the repulsive force computation based on Eq.~\ref{eqn:frderivativesigmoid} accesses the embeddings of a subset of non-adjacent vertices ($v_2$ and $v_5$ in Fig.~\ref{fig:gradla}). 
This repulsive force computation can also be mapped to a general SpMM operation. 


\begin{figure}[!t]
    \centering
    \fbox{\includegraphics[width=0.95\linewidth]{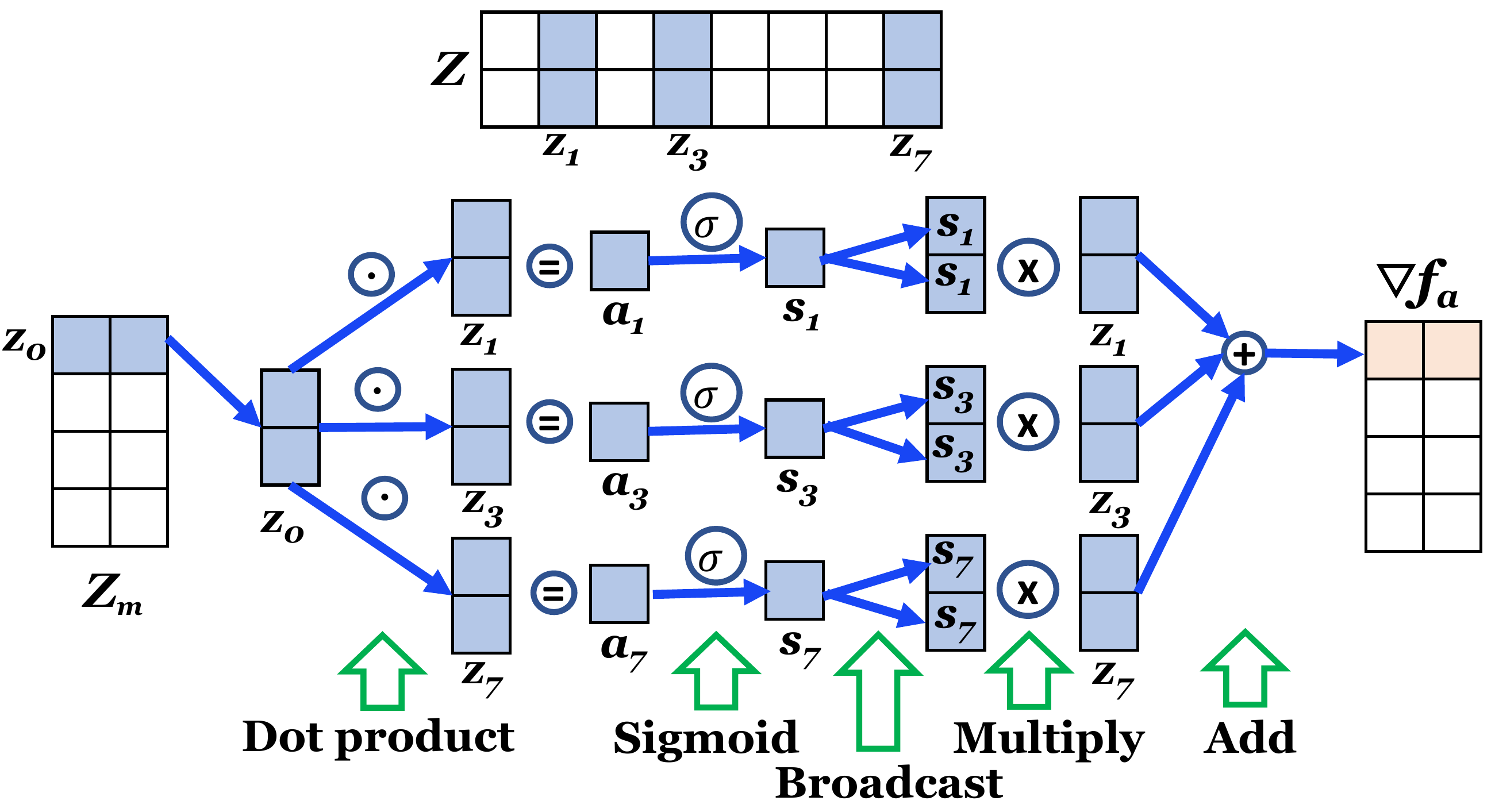}}
    \caption{This example shows gradient computation of attractive force for a single vertex using the sigmoid function.}
    \vspace{-0.5cm}
    \label{fig:simdfig}
\end{figure}

%
While Fig.~\ref{fig:gradla} provides a high-level concept of using SpMM in gradient computation, we cannot directly use an off-the-shelf implementation of SpMM because of the delicacy of various force models.
For example, Fig.~\ref{fig:simdfig} shows how we actually compute $\nabla f_a(v_0)$ based on Eq.~\ref{eqn:faderivativesigmoid}.
At first, we compute dot products between $\vect{z}_0$ and $\vect{z}_1$, $\vect{z}_3$, and $\vect{z}_7$.
These dot products produce scalar values denoted by $a_1$, $a_2$, and $a_7$ that are passed through a sigmoid function generating $s_1$, $s_2$, and $s_7$.
We broadcast these scalar values to form vectors by copying it in each lanes of the vectors that are element-wise multiplied with their corresponding embedding vectors.
The resultant vectors are summed up to finally output $\nabla f_a(v_0)$.
While this computation is conceptually similar to SpMM, the use of sigmoid makes it harder to use standard libraries. 
This mapping allows us to optimally parallelize gradient computation as described below.

{\bf Thread-level parallelization.}
All vertices in a minibatch are independent and compute gradients in parallel. 
For example, different threads can parallelly compute gradients for four vertices (vertices corresponding to $\vect{Z}_b$) in Fig. \ref{fig:gradla}. 
Hence, each thread can sequentially compute $\nabla f_a(v)$ and $\nabla f_r(v)$ following the flow shown in Fig.~\ref{fig:simdfig}. 
When vertices in a minibatch have similar number of neighbors, we use static scheduling where  each thread processes equal number of vertices from the minibatch. 
However, if a vertex has very high degree (commonly seen in scale-free networks), the computation on a high-degree vertex can become the bottleneck.
To address this issue, we compute the degree distribution of a minibatch and partition the minibatch with respect to the number of threads such that each thread processes almost equal number of neighbors.
The latter approach results in a superior load-balancing for scale-free graphs.

{\bf SIMD vectorization.}
While each thread computes $\nabla f_a(v)$ and $\nabla f_r(v)$ in a core, these computations can be further optimized by employing Single Instruction Multiple Data (SIMD) parallelism avaiable in each core.
To this end, we implemented the computations in Fig.~\ref{fig:simdfig} using hardware intrinsics.
Our library includes a code generator that  generates intrinsic codes for different hardware architectures \RevisedText{(similar as~\cite{Whaley2001atlas})}.
In our implementation, we employed extensive register blocking to perform all intermediate computations for a vertex on SIMD registers reducing the write accesses on the output. 
Note that we need to use the full capacity of the SIMD registers of any hardware architecture to effectively implement this technique and general-purpose compilers often are not very good at this level of auto-vectorization. 

\section{Experiments}
\subsection{Experiment setup}
{\bf Experiment overview.}
We conduct three types of experiments to show the effectiveness of \toolname{}: (a) runtime and scalability experiments show that \toolname{} is an order of magnitide faster than other methods considered, (b) visualization experiments show that \toolname{} generates qualitatively and quantitatively  better visualizations, and (c) node classification, link prediction and modularity experiments show that \toolname{} performs similar or better than previous graph embedding algorithms.
We also \RevisedText{provide} insights behind the observed performance of \toolname{}.

{\bf Experiment platform.} We implemented \toolname{} in C++ with multithreading support from OpenMP. 
We used intrinsic functions to exploit SIMD vectorization available on a single core. 
All our experiments were conducted on a server with a dual-socket Intel Xeon Platinum 8160 processors (2.10GHz). The server has 256GB memory, 48 cores (24 cores/socket), and 32MB L3 cache/socket.

\begin{table}[!t]
\centering
\caption{Parameter settings for other methods.}
\begin{tabular}{|c|p{6.2cm}|}
\hline
\textbf{Method} & \textbf{Values of parameters} \\ \hline
DeepWalk~\cite{perozzi2014deepwalk} & walk length = $80$, num. of walks per vertex = $10$ \\ \hline
struc2vec~\cite{ribeiro2017struc2vec} & num. of walks = 20, walk length = 80, window size = 5, num. of layers = 6 and use all optimization options \\ \hline
Verse~\cite{tsitsulin2018verse} & num. of negative samples = $5$ \\ \hline
HARP~\cite{chen2018harp} & \emph{line} model and window size = $2$ \\ \hline
\end{tabular}
\label{tab:parameters}
\vspace{-0.62cm}
\end{table}

{\bf Algorithm settings.}
We used three variants of \toolname{} in \RevisedText{different} experiments: Algorithm \ref{algo:nsforce2vecalgo} with sigmoid function ({\em s\toolname{}}), Algorithm \ref{algo:nsforce2vecalgo} with $t$-distribution ({\em t\toolname{}}), and Algorithms \ref{algo:nsforce2vecalgo} and \ref{algo:nsforce2vecalgorw} with sigmoid function ({\em r\toolname{}}). 
We empirically set the minibatch size $b$ to 384 so that each thread gets $384/48=8$ vertices within a minibatch.
Changing the minibatch size slightly does not impact the convergence. 
In all experiments, we set the number of epochs to 1200, the number of negative samples to 6, and the learning rate to 0.02. We run all experiments 10 times and report average results for all performance measures.
Some of these hyper-parameters (e.g., $\eta$) are obtained by using a grid search discussed in Section \ref{sec:parametersense}.
As Algorithm \ref{algo:nsforce2vecalgorw} performs better for heterogeneous networks, we report results of Algorithm \ref{algo:nsforce2vecalgorw} for multi-label classification.  
We compared \toolname{} with four other embedding methods: DeepWalk~\cite{perozzi2014deepwalk}, HARP~\cite{chen2018harp}, struct2vec~\cite{ribeiro2017struc2vec}, and Verse~\cite{tsitsulin2018verse}. 
We selected these methods because they have different underlying models and support multi-threading. 
Table \ref{tab:parameters} reports various parameters used with these methods. Unless otherwise mentioned explicitly, we generate 128-dimensional embedding and set the number of workers/threads to 48. \RevisedText{In all of our experiments, we use default values for other parameters that have not been stated directly}.

{\bf Datasets for experiments.}
Table~\ref{tab:dataset} reports a set of graphs of various sizes used in our experiments. All these graphs have multiple labels. Most of these graphs have been widely used in previous studies~\cite{perozzi2014deepwalk,tsitsulin2018verse,kipf2016semi,zeng2019graphsaint}. 
 These graphs include networks from scientific publication domain (Cora, Citeseer, Pubmed) and social photo/video sharing networks (Flickr and Youtube). 
 To conduct experiments with large graphs, we have used a social interaction network (Orkut), and a web-crawler network (uk-2005). 
 Note that uk-2005 is the biggest graph that we can solve with 256 GB memory in our server. 
 The ground truth node labels are not available for Orkut and uk-2005 graphs. Thus we perform only link prediction task for these graphs. Cora, Citeseer, Pubmed and Flickr networks are homogeneous i.e., each node is assigned to a single label. Thus node classification task on those networks automatically becomes a multi-class classification problem. On the other hand, Youtube graph is heterogeneous where one node can have more than one labels. So, node classification on Youtube is a multi-label classification problem.


\begin{table}[!t]
\centering
\caption{Datasets used for graph embedding experiments. Graphs are available at \url{https://linqs.soe.ucsc.edu/data} and \url{https://sparse.tamu.edu/} [Last accessed: September 4, 2020].}
\begin{tabular}{|c|c|c|c|c|}
\hline
\textbf{Graphs} & \textbf{Vertices} & \textbf{Edges} & \textbf{\#Labels} & \textbf{Avg. Degree} \\ \hline
Cora            & 2,708              & 5,429           & 7                 & 3.89                 \\ \hline
Citeseer        & 3,327              & 4,732           & 6                 & 2.736                \\ \hline
Pubmed  &   19,717	&   44,338	&   3	&   4.49  \\ \hline
Flickr          & 89,250	    &   899,756	&   7	&   20.16              \\ \hline
Youtube         & 1,138,499           & 2,990,443        & 47                & 5.253                \\ \hline
Orkut         & 3,072,441 & 117,185,083         & -                & 76.28                \\ \hline
uk-2005         & 39,459,925  & 936,364,282          & -                & 47.46                \\
\hline
\end{tabular}
\label{tab:dataset}
\vspace{-0.64cm}
\end{table}
\begin{table*}[!t]
\centering
\caption{Time (sec.) taken by different tools to generate embeddings using 48 threads. t\toolname{}, s\toolname{}, and r\toolname{} represent Algorithm \ref{algo:nsforce2vecalgo} with $t$-distribution, Algorithm \ref{algo:nsforce2vecalgo} with sigmoid function, and Algorithms \ref{algo:nsforce2vecalgo} with random walks from Algorithms \ref{algo:nsforce2vecalgorw} and sigmoid function, respectively. Numbers in the parenthesis represents speedups of t\toolname{}.}
\begin{tabular}{|l|c|c|c|c|c|c|c|}
\hline
\textbf{Methods}  & {\color[HTML]{663234} \textbf{DeepWalk}} & {\color[HTML]{1126FF} \textbf{HARP}} & {\color[HTML]{F8A102} \textbf{struc2vec}} & {\color[HTML]{00D2CB} \textbf{Verse}} & {\color[HTML]{009901} \textbf{tForce2Vec}} & {\color[HTML]{009901} \textbf{sForce2Vec}} & {\color[HTML]{009901} \textbf{rForce2Vec}} \\ \hline
\textbf{Cora}     & 73.95 ($\approx$52$\times$)                     & 8.40 ($\approx$6$\times$)             & 274.52 ($\approx$192$\times$)                    & 46.99 ($\approx$33$\times$)                  & 1.43   &   1.35    &   1.47                                         \\ \hline
\textbf{Citeseer} & 87.24 ($\approx$51$\times$)                     & 5.75 ($\approx$3.4$\times$)             & 367.05 ($\approx$214$\times$)                    & 55.94 ($\approx$33$\times$)                  & 1.71                                       & 1.46                                       & 1.69                                          \\ \hline
\textbf{Pubmed}   & 591.48 ($\approx$56$\times$)                    & 98.61 ($\approx$9.4$\times$)               & 2199.16 ($\approx$209$\times$)                   & 300.35 ($\approx$28.6$\times$)                 & 10.49                                      & 9.36                                      & 10.37                                          \\ \hline
\textbf{Flickr7}  & 1989.56 ($\approx$37$\times$)                   & 1772.73 ($\approx$33$\times$)             & 8176.75 ($\approx$151.5$\times$)                   & 1297.92 ($\approx$24$\times$)                & 53.98                                      & 47.48                                      & 65.08                                         \\ \hline
\textbf{Youtube}  & 28049.52 ($\approx$42.3$\times$)                  & 10854.49 ($\approx$16.4$\times$)              & x                                         & 15798.77 ($\approx$24$\times$)               & 662.81                                     & 654.75                                        & 624.73                                          \\ \hline
\textbf{Orkut}    & 86548.58 ($\approx22\times$)                  & x              & x                                         & 41624.98 ($\approx$10.5$\times$)               & 3952.93                                    & 3311.97                                   & 3901.16                                          \\ \hline
\textcolor{black}{\textbf{uk-2005}}    & x                  & x              & x                                         & 413375.62 ($\approx45.4\times$)               & 9114.93                                    & 9253.8                                  & 14900.9                                         \\ \hline
\end{tabular}
\label{tab:runtimes}
\vspace{-0.5cm}
\end{table*}
\vspace{-0.25cm}
\subsection{Runtime and scalability experiments}
\vspace{-0.05cm}
{\bf Runtime on 48 threads.}
Table \ref{tab:runtimes} reports the total runtime using 48 threads for different methods to generate 128-dimensional embeddings. 
The numbers within the parentheses of other methods denote the speed-up of t\toolname{} over the corresponding method. 
Overall, \toolname{} is on average 43$\times$ and 28$\times$ faster than DeepWalk and Verse, respectively.
\toolname{} is more than 100$\times$ faster than struc2vec and more than 10$\times$ faster than HARP for large graphs. 
We observe that HARP is usually fast for smaller graphs, but it becomes increasingly slower as the graph becomes bigger. 
We also found that DeepWalk, struc2vec and HARP cannot process larger graphs because of their high computational and/or memory requirement.

{\bf Runtime for a billion-edge graph.}
To conduct experiments with a very large graph, we select uk-2005 web-crawler graph with 39 million vertices and nearly one billion edges. 
For uk-2005, we generated 64-dimensional embedding so that it can be stored in the memory.
DeepWalk, HARP, and struc2vec fail for uk-2005 due to memory error. 
Verse takes nearly 5 days (413375.62 seconds) to generate embedding. 
By contrast, s\toolname{} takes 2.57 hours to generate embedding, which is around 45$\times$ faster than Verse.
Thus, \toolname{} advances the state-of-the-art in big graph analytics by enabling the embedding of large-scale graphs that are becoming more and more common in various science and business applications.

\begin{figure}[!t]
    \centering
    \includegraphics[width=0.49\linewidth,height=3.6cm]{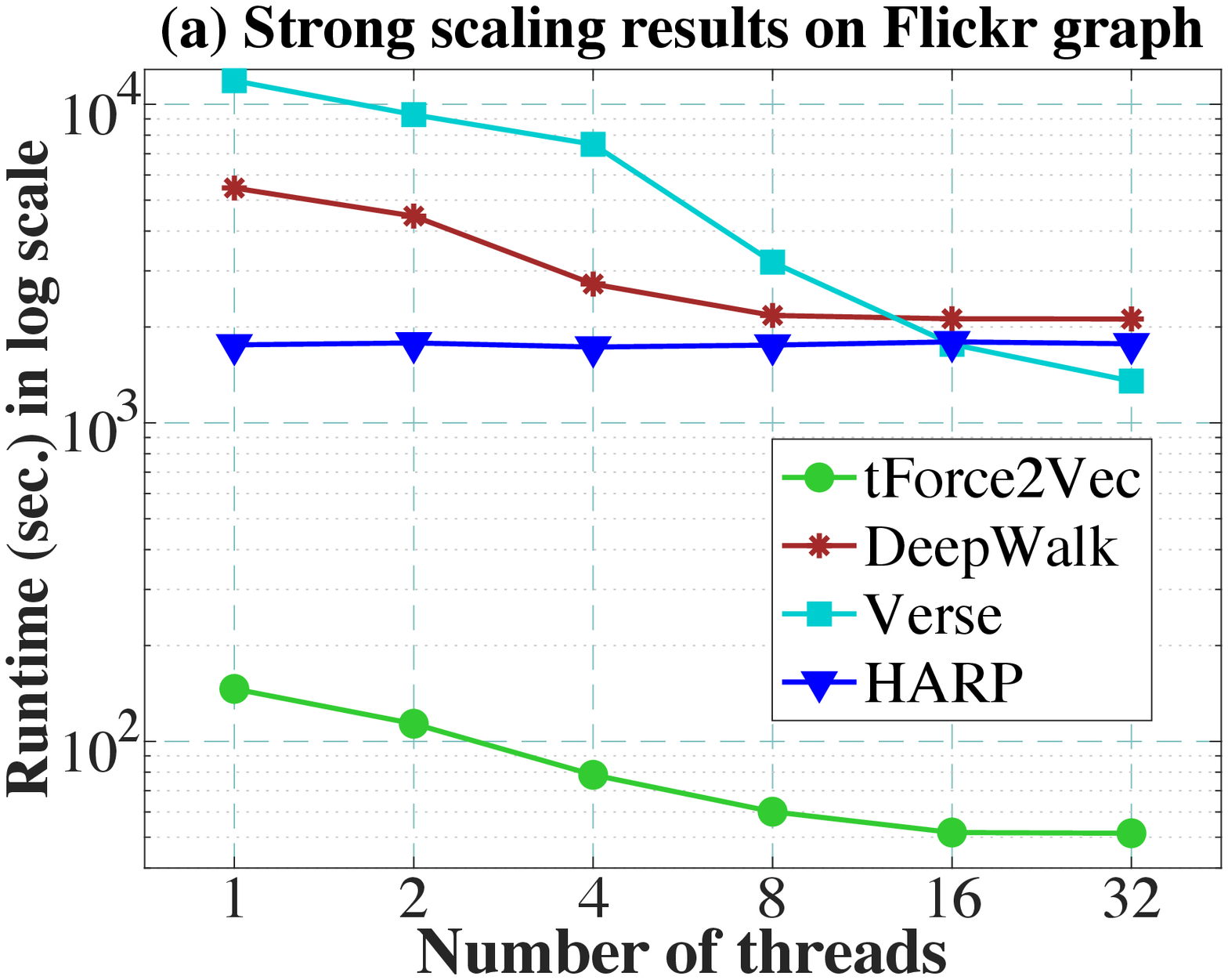}
    \includegraphics[width=0.49\linewidth,height=3.6cm]{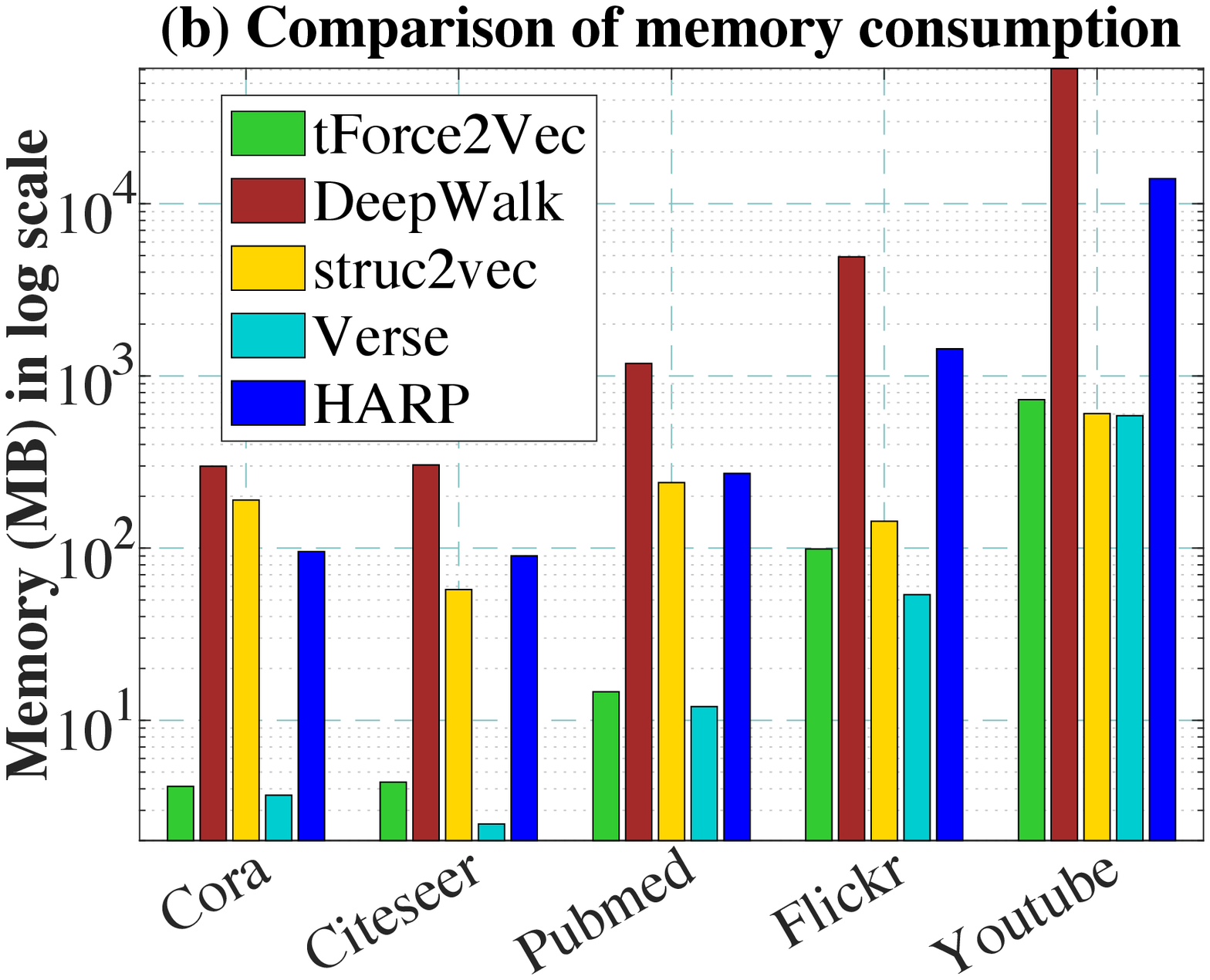}
    \caption{(a) Strong scaling results for the Flickr graph with different number of threads. (b) Memory consumption by different methods for five different benchmark datasets. }
    \label{fig:runtimememory}
    \vspace{-0.6cm}
\end{figure}
{\bf Scalability.}
Fig.~\ref{fig:runtimememory}(a) shows the strong scaling results for the Flickr graph (the biggest graph where all methods are successful). 
We observe that both t\toolname{} and Verse scales well as we increase the number of threads.
However, t\toolname{} runs an order of magnitude faster on all thread counts. 
The HARP code obtained from the repository provided by authors does not scale well.

{\bf Why does \toolname{} perform so well?}
The faster runtime of \toolname{} originates from three aspects of our algorithm and implementation.
First, the mapping of core gradient computation to linear algebra operation shown in Fig.~\ref{fig:gradla} helps us 
simplify and parallelize the core computations which dominate the runtime of the algorithm.
We completely eliminated \emph{synchronization} %
and \emph{false sharing} that can happen due to asynchronous or dependent read-write operations in memory \cite{bolosky1993false}.
We avoid this dependency by employing batch processing of vertices.
Second, we fully utilized SIMD parallelization using 
low-level intrinsics codes available for modern processors %
to get the full capacity of the hardware.
Our implementation provides up to 2.7$\times$ speedup over the implementation which does not use intrinsic and solely relies on the auto-vectorization of compiler. 
Third, \toolname{} utilizes cache hierarchies as much as possible. 
Our linear algebra kernels stream data to processor caches and registers and utilize in-cache data as much as possible.  In our intrinsic implementation, we also use extensive register-blocking to eliminate the intermediate write accesses of the output data. 
Hence, we are able to eliminate unnecessary data accesses.
For example, the L1 cache miss rate in \toolname{} is never greater than 2\%, whereas the L1 cache miss rate can be as high as 16.5\% in Verse. 
Since memory accesses (in terms of memory bandwidth and latency) are the primary bottleneck in graph analysis, \toolname{} performs well by utilizing spatial and temporal data locality.

{\bf Memory requirements.}
Fig.~\ref{fig:runtimememory}(b) shows the memory consumption by different methods. 
\toolname{} stores the graph in the Compressed Sparse Row (CSR) format and uses 4 bytes for vertex indices and edge weights. 
We also use single-precision floating point numbers to store 128-dimensional embedding.
Thus, the estimated memory cost for \toolname{} is $4n+8m+512n$ bytes or $516n+8m$ bytes. 
We observe that Verse and t\toolname{} consume less memory than DeepWalk and HARP. 
For example, in Fig.~\ref{fig:runtimememory} (b), DeepWalk, HARP, Verse and tForce2Vec empirically consume peak memory of around 60 GB, 13.4 GB, 587 MB, and 727 MB for the Youtube graph, respectively. 

\begin{figure*}[!t]
    \centering
    \fbox{\includegraphics[width=0.23\linewidth,height=3cm]{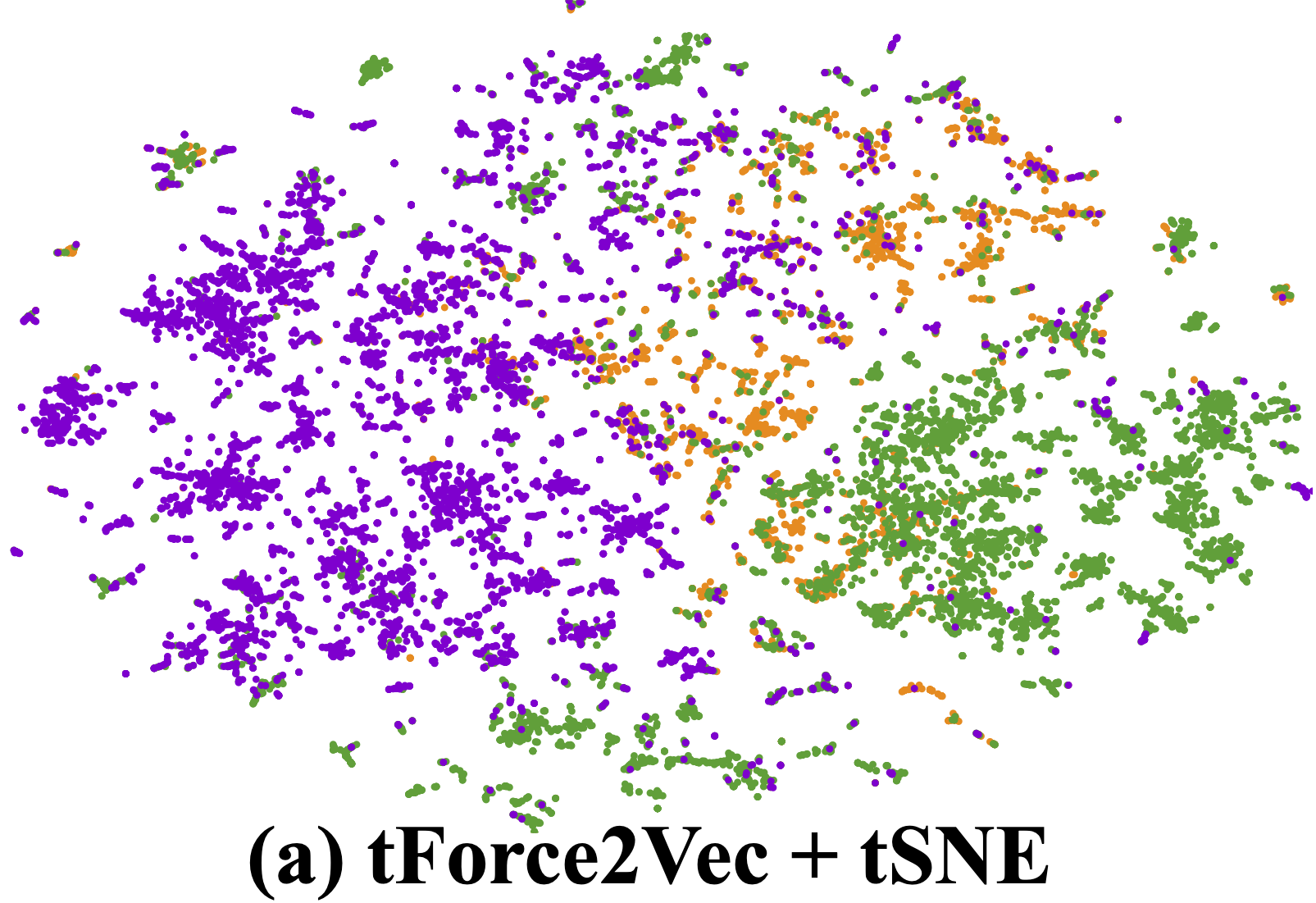}}
    \fbox{\includegraphics[width=0.23\linewidth,height=3cm]{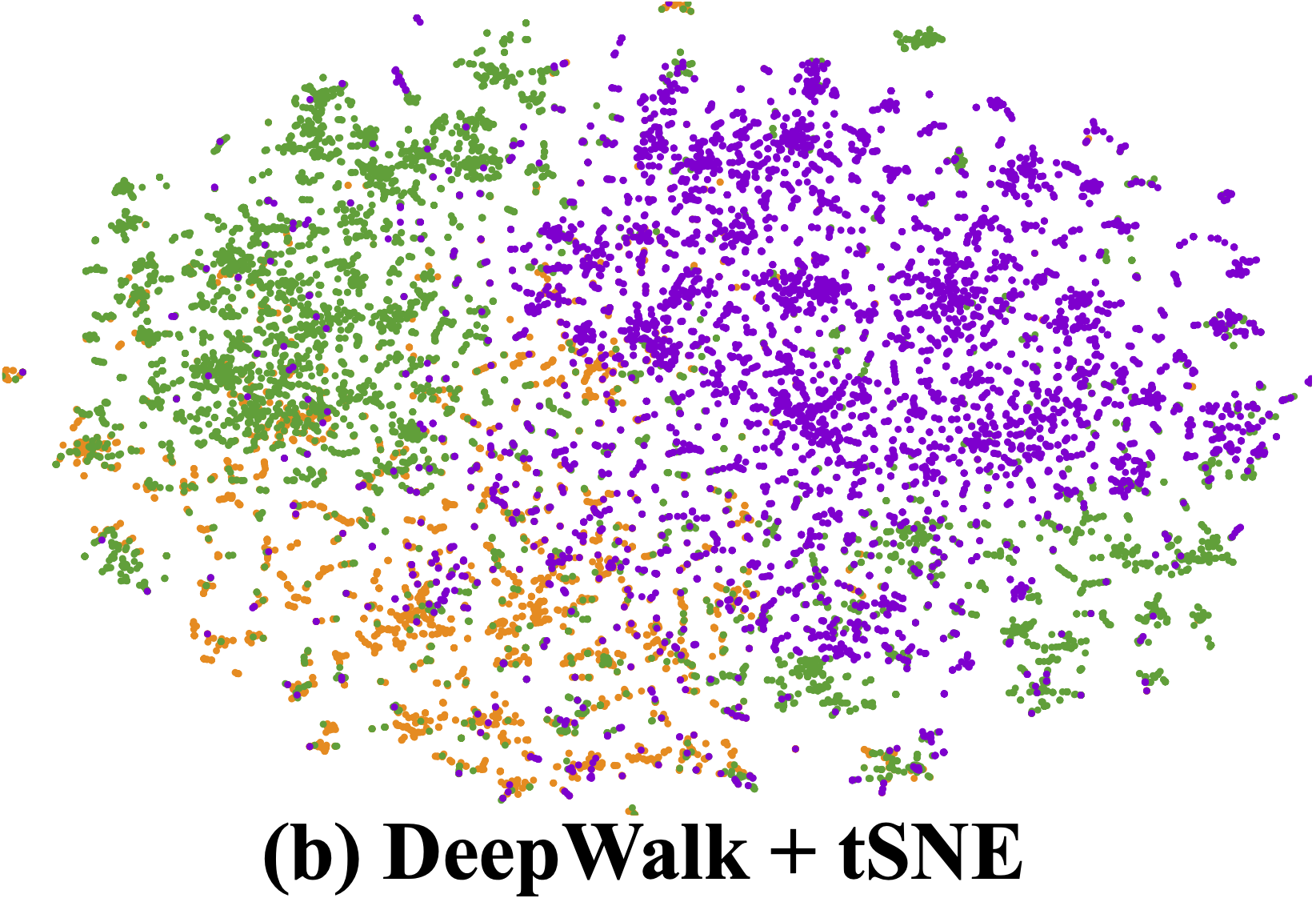}}
    \fbox{\includegraphics[width=0.23\linewidth,height=3cm]{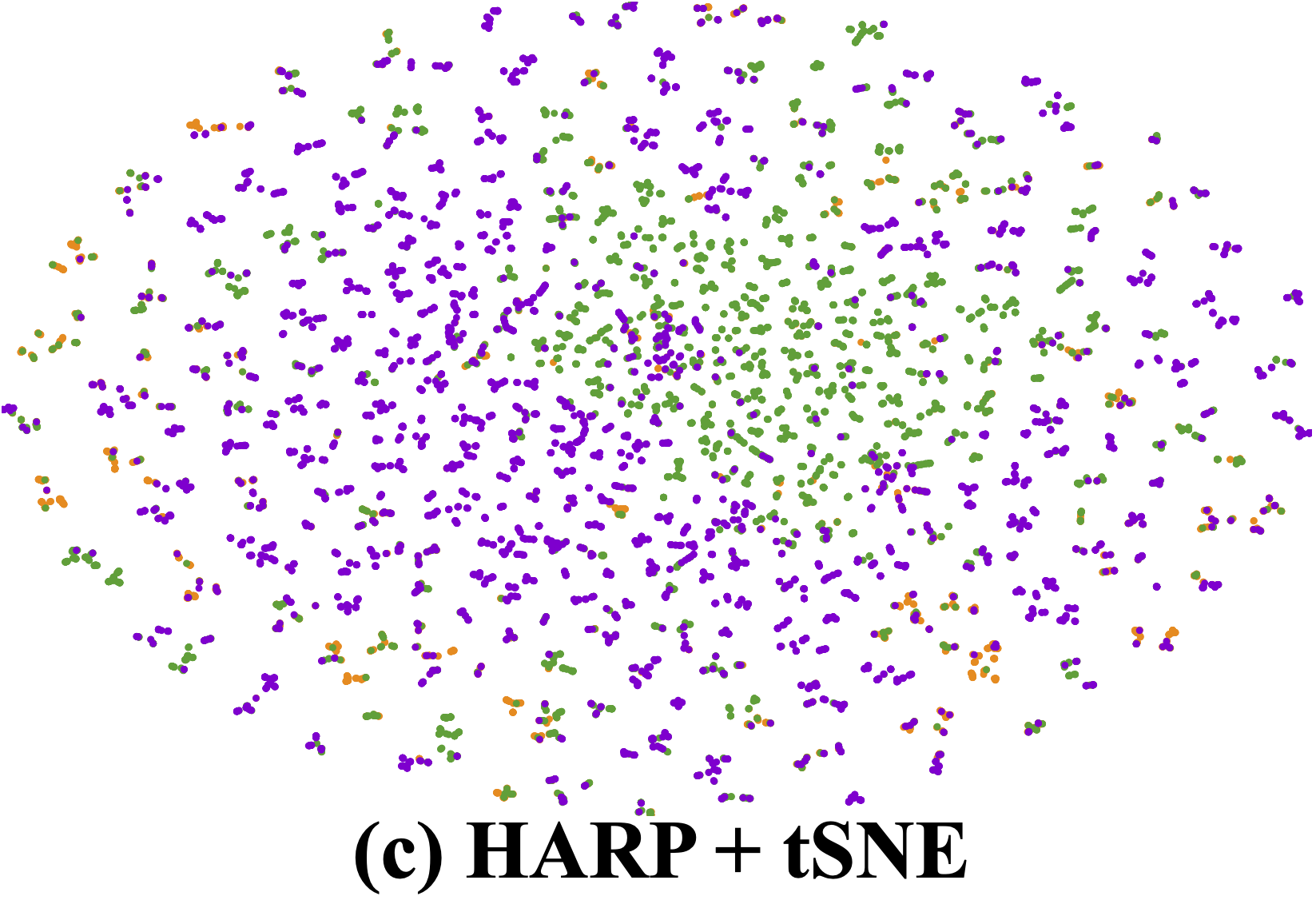}}
    \fbox{\includegraphics[width=0.23\linewidth,height=3cm]{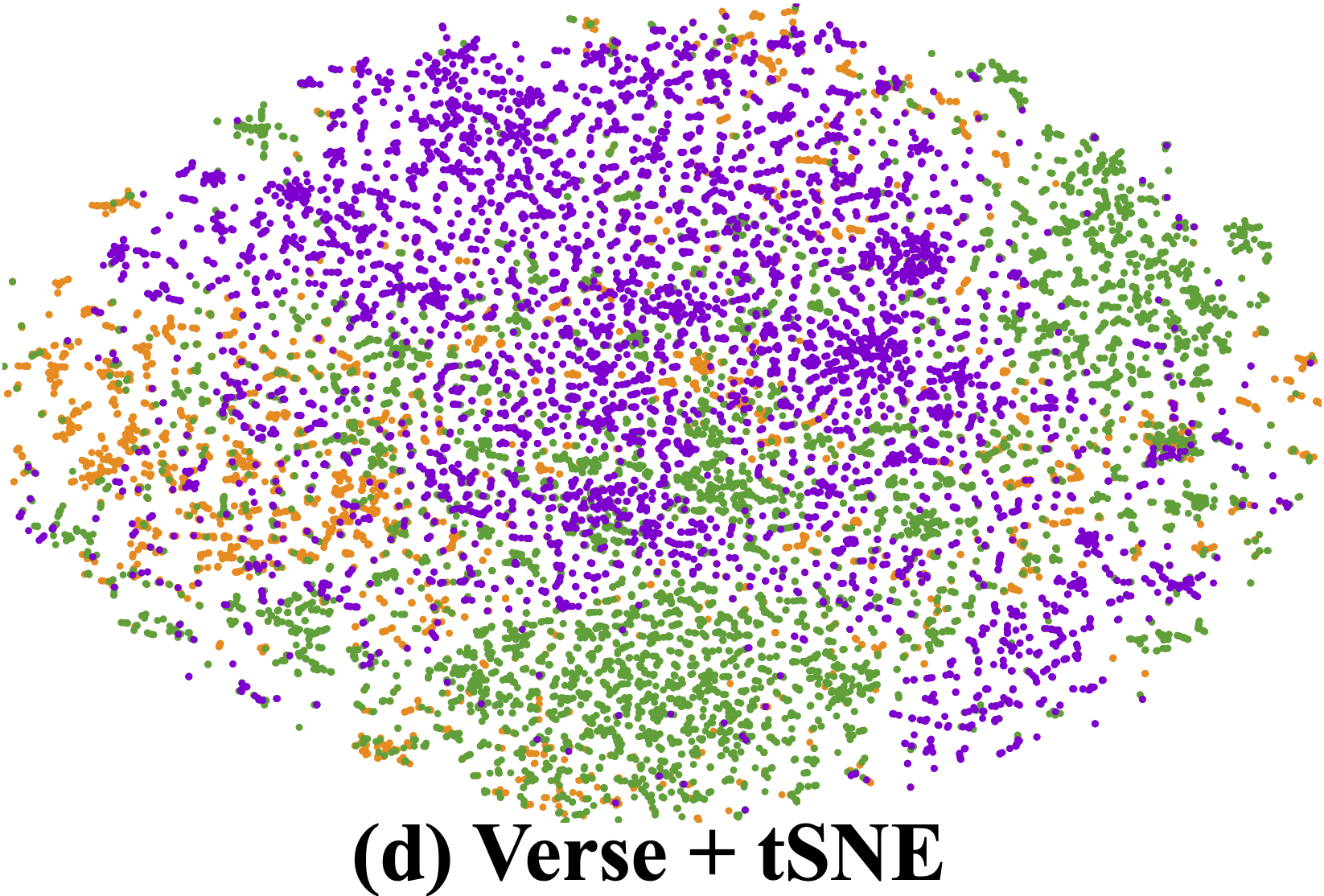}}
    \caption{Visualization of 2D projections for Pubmed dataset by t-SNE from 128 dimensional embeddings generated by (a) tForce2Vec, (b) DeepWalk, (c) HARP, and (d) Verse. Colors represent respective classes in the dataset.}
    \label{fig:abcdvis}
    \vspace{-0.5cm}
\end{figure*}

\begin{figure}[!t]
    \centering
    \fbox{\includegraphics[width=0.44\linewidth,height=2.8cm]{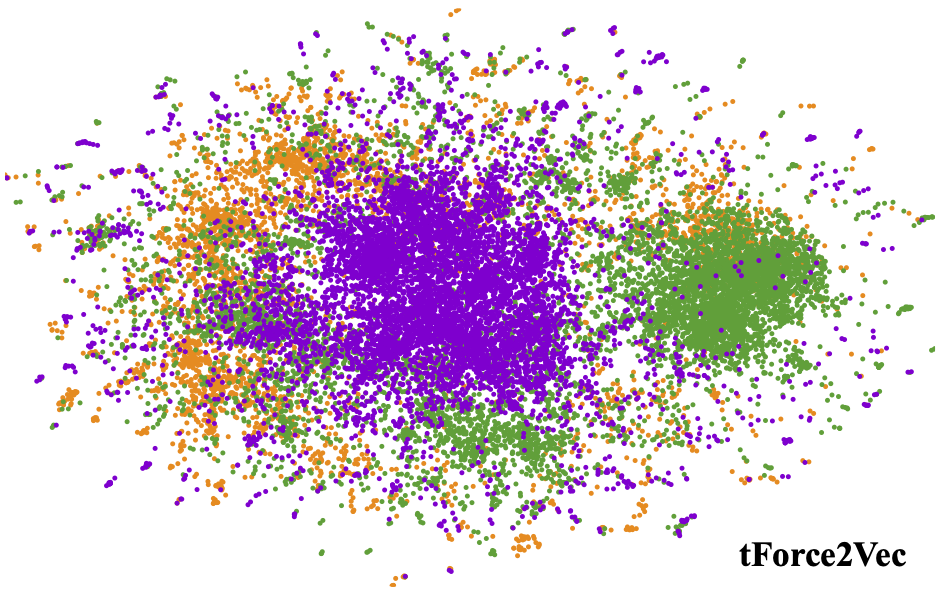}}
    \fbox{\includegraphics[width=0.44\linewidth,height=2.8cm]{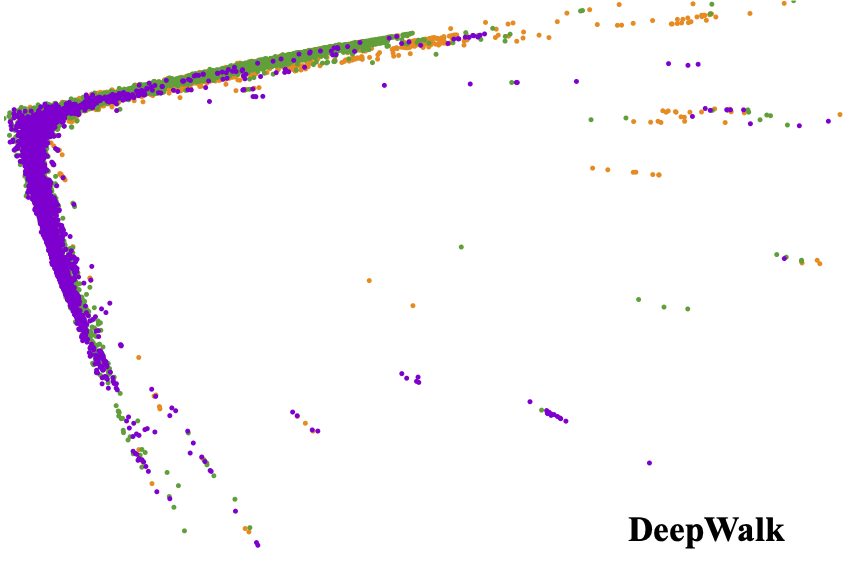}}
    \caption{Visualization of 2D embeddings for Pubmed dataset generated by tForce2Vec (left) and DeepWalk (right). Colors represent respective classes in the dataset.}
    \label{fig:2dvis}
    \vspace{-0.3cm}
\end{figure}

\subsection{Visualization quality}
A primary motivation of this work is to generate an embedding that leads us to an aesthetically pleasing visualization.
Graph visualization is especially important for exploratory science and interpretation of results. 
Here, we evaluate visualization quality both qualitatively and quantitatively using 128D and 2D embeddings. 

{\bf Visualizing 128D embedding using t-SNE.}
To visualize graphs, we use t-SNE~\cite{maaten2008visualizing} to generate 2D projections from  128D embeddings obtained by various embedding algorithms. 
Fig.~\ref{fig:abcdvis} shows visualizations of the Pubmed graph, where different colors denote different classes of vertices.
A good embedding should group vertices of the same class together to form a cluster while separating one class from others as much as possible. 
Fig.~\ref{fig:abcdvis} shows that t\toolname{} preserves the class structures better than other methods as it can make different classes well separated.
The next best visualization is provided by DeepWalk. 
Fig.~\ref{fig:abcdvis} shows that most methods are able to form small clusters from vertices of a particular class; that is, they preserve local structures of clusters reasonably well.
However, t\toolname{} and to some extent DeepWalk preserve the global structure better than other methods. 

{\bf Quantitative evaluation of the visualization.}
To evaluate these visualizations quantitatively, we use $Q_{local}$ and $Q_{global}$ quality measures \cite{lee2010scale} for all lower dimensional projections by t-SNE using $\mathbf{dimRed}$ R package. $Q_{local}$ reflects the preservation of local structure in clusters i.e., how one cluster of a class is separated from other clusters of classes. $Q_{global}$ represents the global structure preservation of clusters where data points of a class are expected to form one cluster. For both of these measures, higher values mean better results. We report the quantitative values in Table \ref{tab:quantitativemeasures}. The quantitative measures also support that t\toolname{} preserves more global structure than other methods. This is due to the robustness and quality of the embedding generated by t\toolname{}. 

\begin{table}[!t]
\centering
\caption{Quantitative measures \cite{lee2010scale} for visualization by t-SNE shown in Fig.~\ref{fig:abcdvis}. Higher value represents better result.}
\begin{tabular}{|c|c|c|c|c|}
\hline
\textbf{Measures} & \textbf{tForce2Vec} & \textbf{Verse} & \textbf{DeepWalk} & \textbf{HARP} \\ \hline
$Q_{local}$ &	0.64	& 0.48 &	0.58 &	0.77 \\ \hline
$Q_{global}$	& 0.23	&   0.04	&   0.07	&   0.05 \\ \hline
\end{tabular}
\label{tab:quantitativemeasures}
\vspace{-0.6cm}
\end{table}

{\bf Visualizations from direct 2D embeddings.}
Algorithm \ref{algo:nsforce2vecalgo} computes \emph{attractive} and \emph{repulsive} forces based on neighbors and non-neighbors, respectively. As mentioned earlier, this force-directed model is widely used to generate 2D or 3D graph layouts.  
Fig.~\ref{fig:2dvis} shows the visualization of the Pubmed graph obtained from direct 2D embeddings from  t\toolname{} and DeepWalk.
While t\toolname{} can still generate a good embedding, DeepWalk fails to visualize a graph (without the help of t-SNE).
Thus, t\toolname{} has a clear advantage in visual graph analytics over other graph embedding methods.


\subsection{The effectiveness of embeddings in ML tasks}
Now we discuss the effectiveness of graph embedding methods in traditional prediction tasks such as node classification, link prediction and community detection.
Specifically, we use $F1$-micro and $F1$-macro scores to assess the quality of link prediction and node classification tasks. To compare clusterings, we use the \emph{Modularity} score. 



    

\begin{figure}[!t]
    \centering
    \includegraphics[width=0.49\linewidth,height=3.6cm]{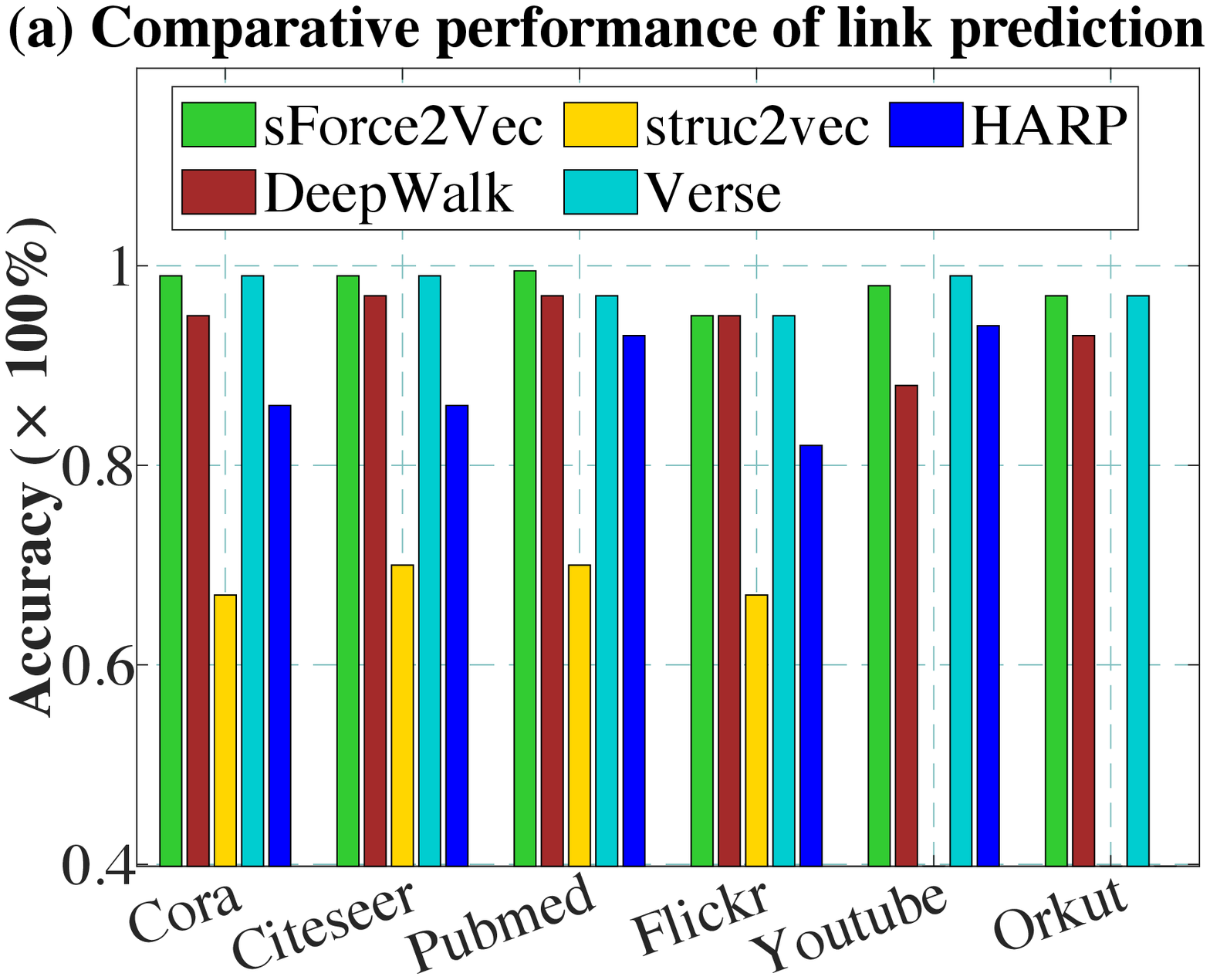}
    \includegraphics[width=0.49\linewidth,height=3.6cm]{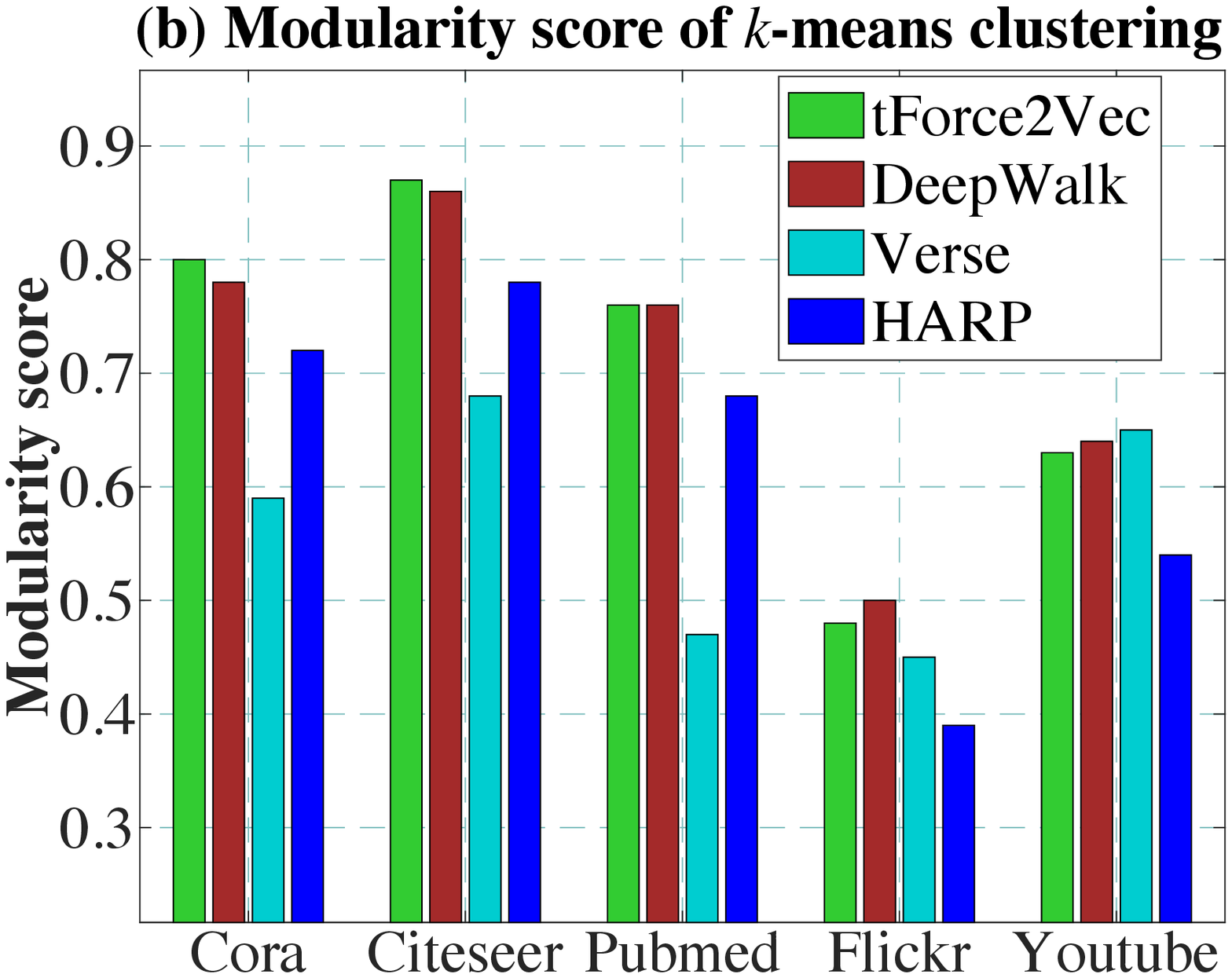}
    \caption{(a) Comparative performance of link prediction task by different methods on different datasets. (b) Comparative results of clustering task by different methods.}
    \label{fig:linkcluster}
    \vspace{-0.75cm}
\end{figure}

\begin{figure*}[!t]
    \centering
    \includegraphics[width=0.245\linewidth,height=3.6cm]{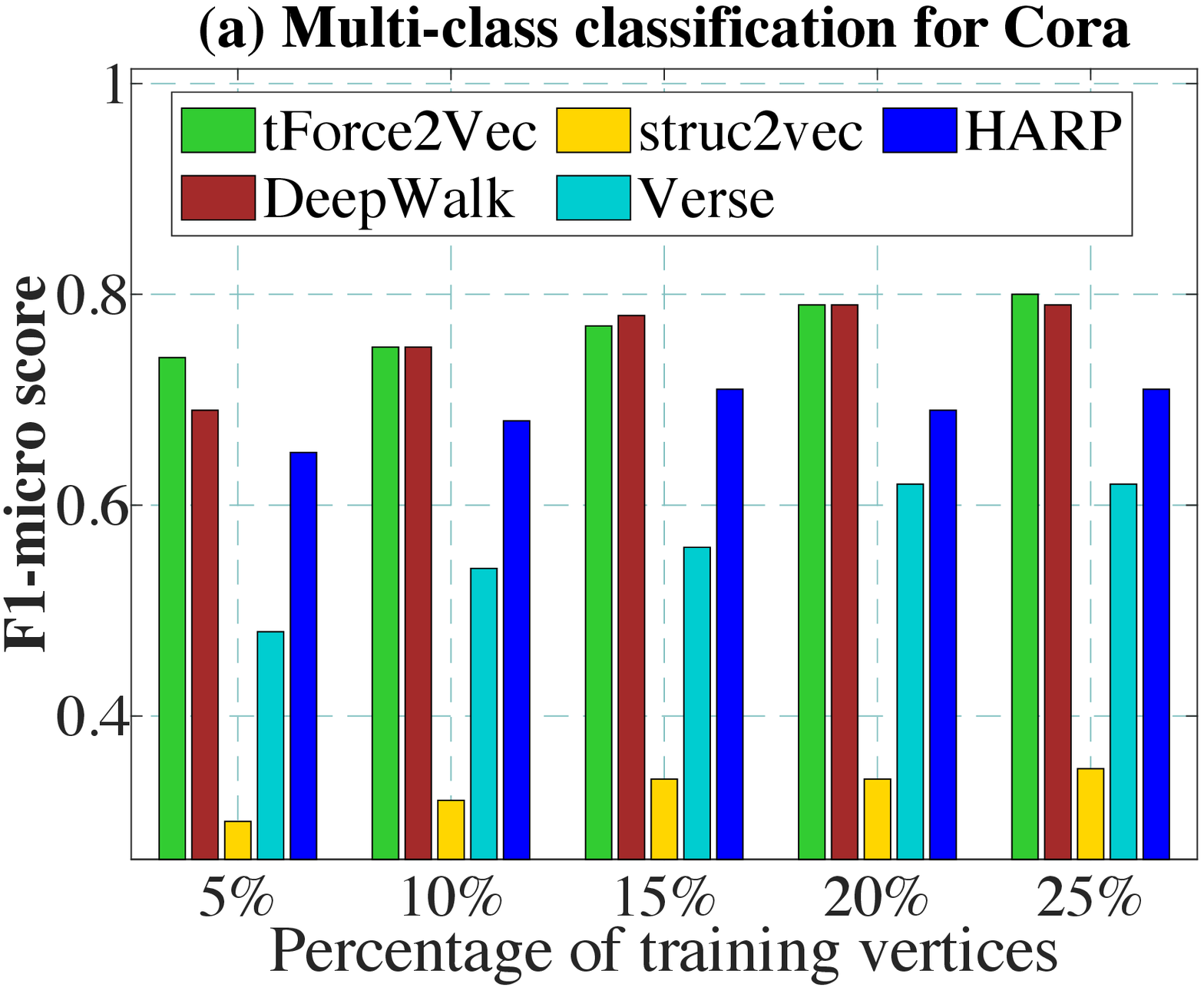}
    \includegraphics[width=0.245\linewidth,height=3.6cm]{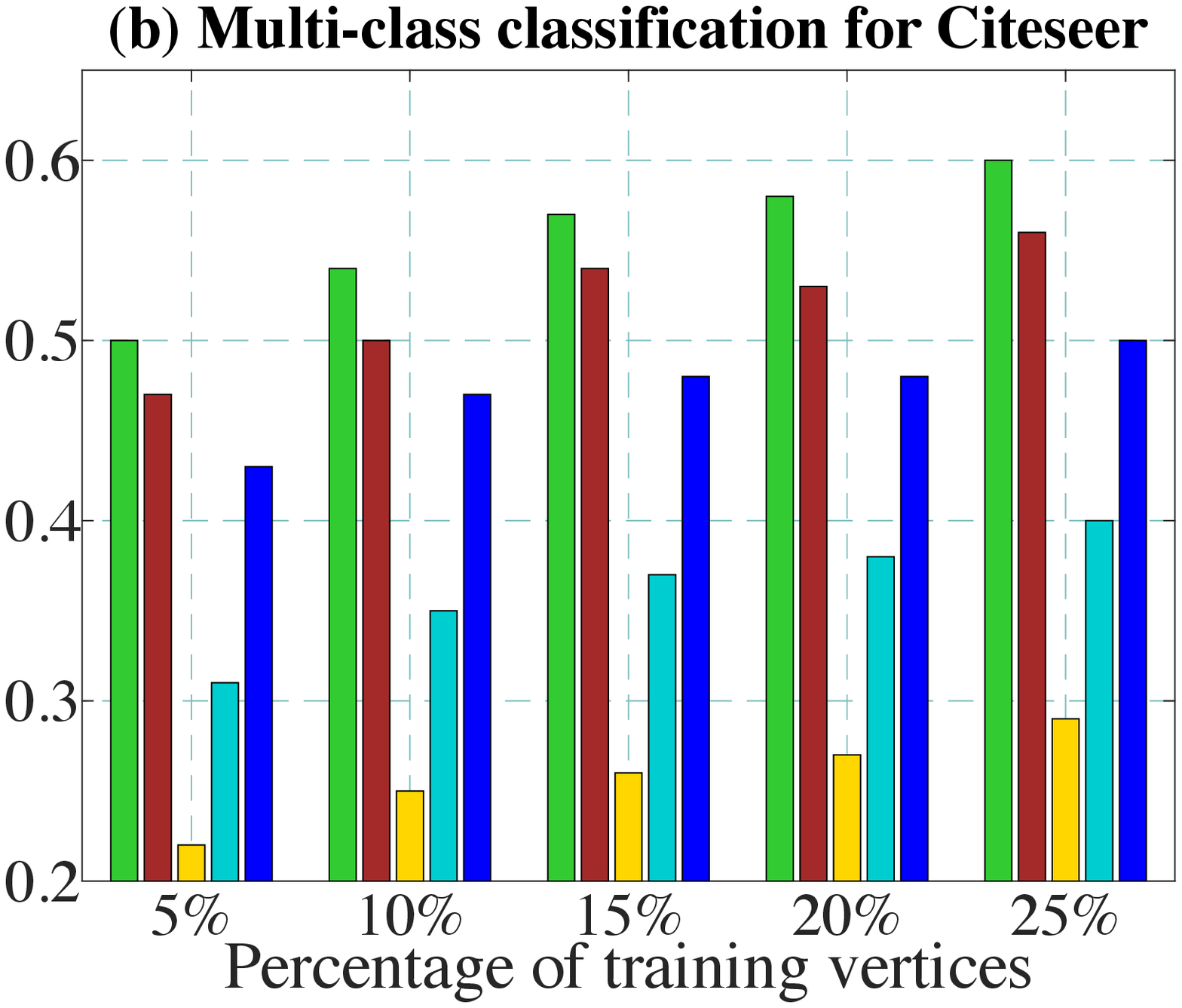}
    \includegraphics[width=0.245\linewidth,height=3.6cm]{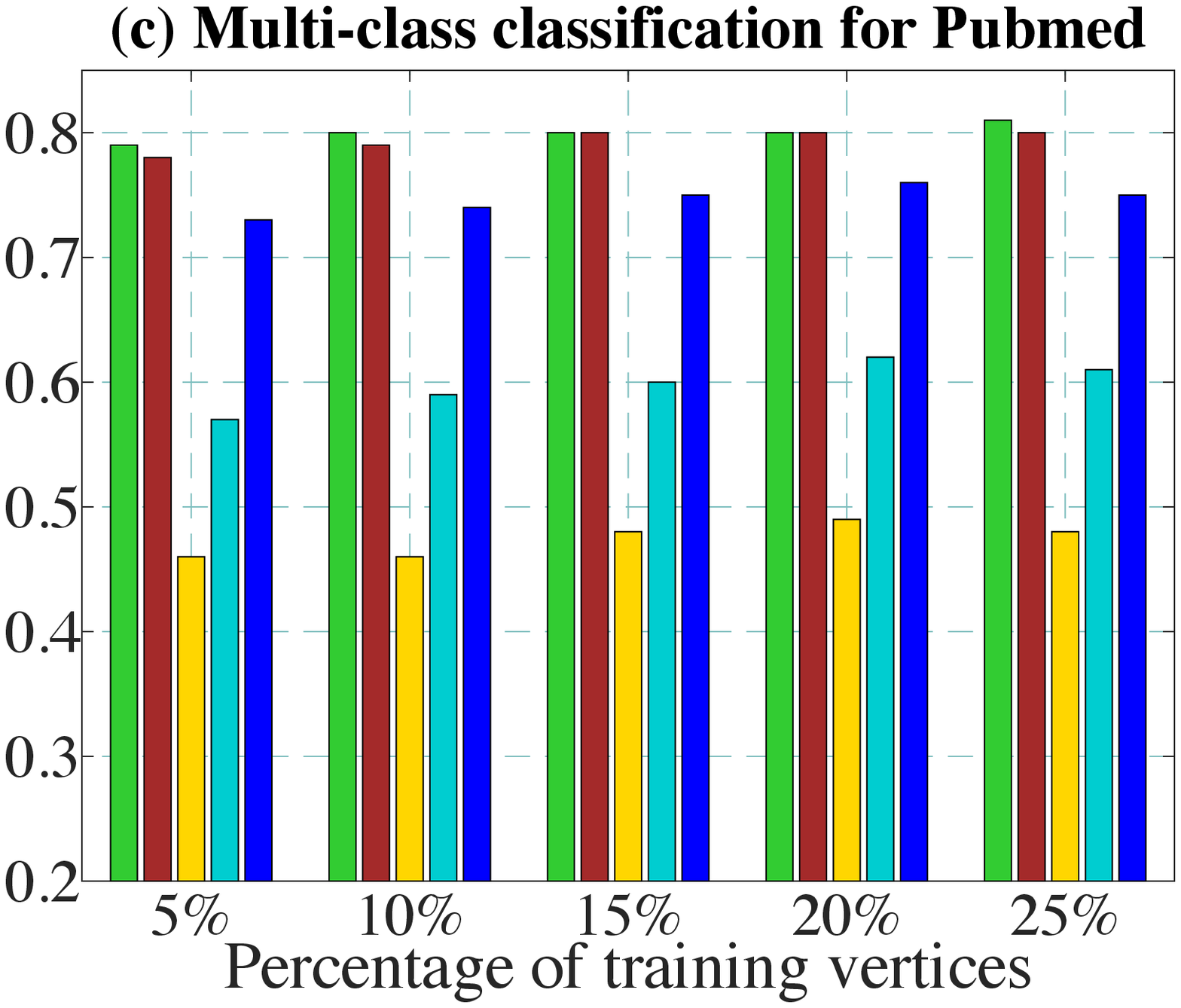}
    \includegraphics[width=0.245\linewidth,height=3.6cm]{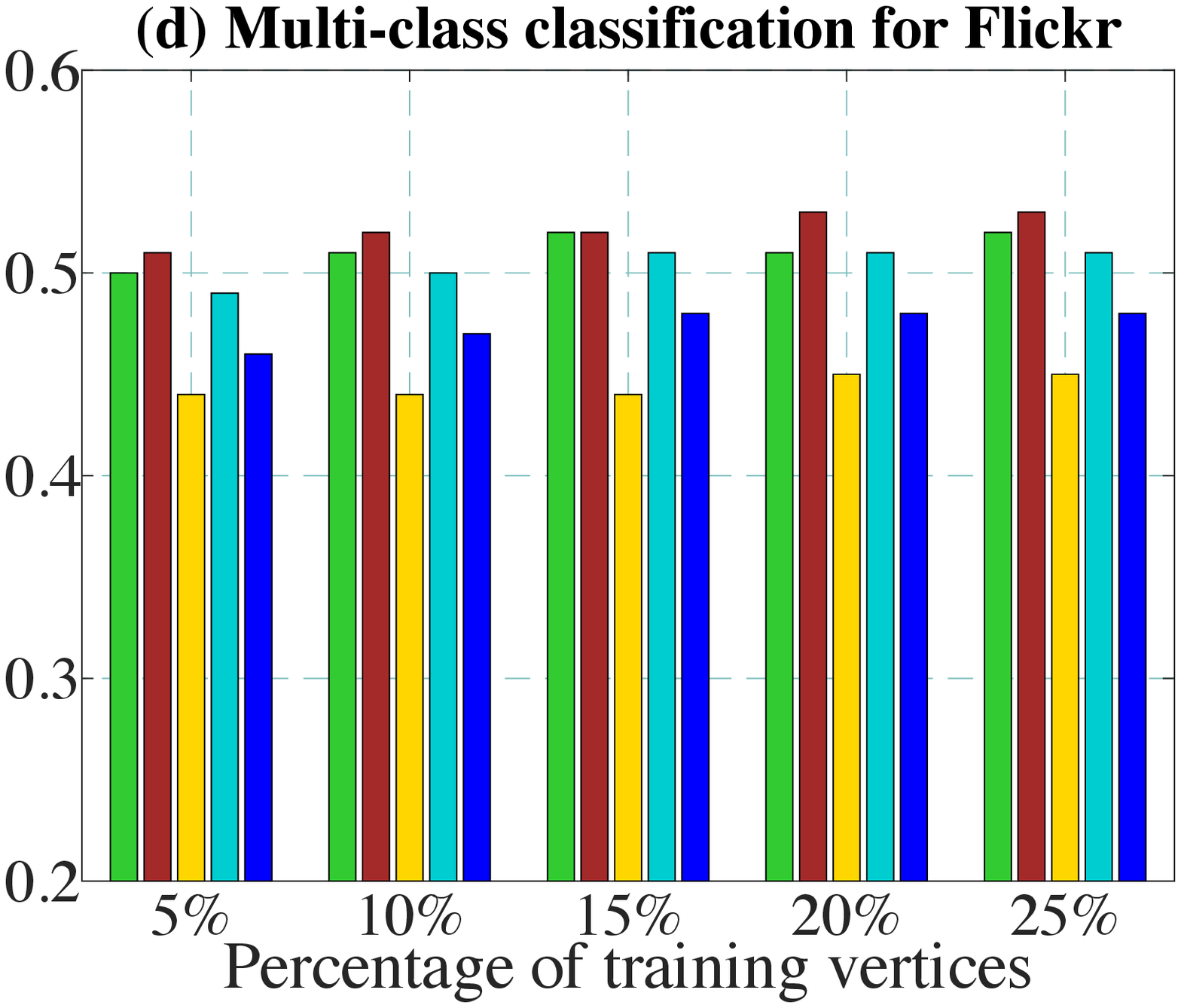}
    \caption{F1-micro scores of node classification task on embeddings generated from (a) Cora, (b) Citeseer, (c) Pubmed, and (d) Flickr datasets by different graph embedding methods.}
    \label{fig:abcclassification}
    \vspace{-0.65cm}
\end{figure*}

{\bf Link prediction.}
To predict links (edges) from the embedding, we reconstruct an edge by performing an element-wise vector operation between the embeddings of a pair of neighboring vertices in the graph.
Following the guidance of prior work~\cite{grover2016node2vec,tsitsulin2018verse}, we use a  dHadamard vector operator that performs element-wise multiplication of $d$-dimensional vectors.
Basically, we treat a pair of adjacent vertices as a positive sample and a pair of non-adjacent vertices as a negative sample.
This formulation converts the link predication task into a binary classification problem. In any type of binary classification, we define accuracy to be $\frac{TP+TN}{TP+TN+FP+FN}$, where TP, FP, TN, and FN represent the absolute number of true positive, false positive, true negative and false negative predictions, respectively. We prepare an evenly distributed dataset having equal number of positive and negative samples. We shuffle them and use 50\% of the samples to train a logistic regression model while rest of the samples for testing. We report the results in Fig.~\ref{fig:linkcluster}(a). We observe that s\toolname{} and Verse perform better than other methods.
DeepWalk also performs generally well, but the embeddings from HARP and struc2vec can not predict links with high accuracy. 
Overall, s\toolname{} takes much less time to generate an embedding that are either better or competitive for the link prediction task compared to other graph embedding algorithms.

For uk-2005 with about 1 billion edges, we could not conduct a link prediction experiment using all edges due to memory limitation. 
Hence, we create an induced subgraph from a randomly selected subset of 1\% vertices and edges adjacent to these sampled vertices.  
Then, we predict links from this induced subgraph using s\toolname{} and Verse (other methods failed to generate embeddings for this graph). We observe that s\toolname{} and Verse achieve accuracy of 95\% and 97\%, respectively. Note that s\toolname{} attains this competitive accuracy significantly faster than Verse.

{\bf Clustering.}
To extract the community structure \RevisedText{from the embedding of a graph}, we follow the technique used in Verse \cite{tsitsulin2018verse} and use the $k$-means algorithm to cluster 128-dimensional embeddings. 
If an embedding captures the community structure effectively, each cluster identified by $k$-means should represent a community in the original graph.  
We evaluate the quality of a $k$-means clustering by the \emph{modularity} score~\cite{blondel2008fast} that  computes the fraction of the edges that are within a given cluster minus the expected fraction, if edges are distributed randomly. 
To identify the best clustering solution, we run $k$-means for the number of clusters from 2 to 50 and report the best clustering solution measured by the modularity score (higher is better).
Fig.~\ref{fig:linkcluster}(b) shows that t\toolname{} achieves the best or near best modularity score.
DeepWalk also performs well for all graphs, but the performance of other methods fluctuates across graphs. 
For example, Verse performs the best for Youtube, but does not perform well for other graphs. 
In summary, t\toolname{} is robust in preserving community structures in the embedding space while providing embeddings much faster than other methods. 


{\bf Node classification.}
In a node classification task, we aim to predict node labels from the embedding of nodes. 
We use a random subset of vertices and their embeddings to train a logistic regression model and then, predict the labels of the rest of the vertices (that are not used in training).
We evaluate the quality of the multi-class classification by the {\em $F1$-micro} score that aggregates the contributions of all classes to calculate the average value of the final $F1$ score. 
For this case, the precision and recall for a set of classes $C$ are defined as follows: $P = \frac{\sum_{c\in C}TP_c}{\sum_{c\in C}(TP_c+FP_c)}$, $R=\frac{\sum_{c\in C}TP_c}{\sum_{c\in C}(TP_c+FN_c)}$, where $TP_c$, $FP_c$, and $FN_c$ denote the number of true positive, false positive, and false negative predictions of class $c$, respectively.
Then, the $F1$-micro score is computed by $\frac{2*P*R}{P+R}$.
    
We train a logistic regression (one vs. rest) model by taking 5\%, 10\%, 15\%, 20\%, and 25\% of the nodes in training set, respectively and report the F1-micro score in Fig.~\ref{fig:abcclassification} by making prediction on rest of the nodes as the testing set. 
We observe that t\toolname{} and DeepWalk generally perform better than other methods.
For Citeseer, t\toolname{} performs much better than other approaches. 
The accuracy of struc2vec and Verse are not competitive, while HARP performs reasonably well.    
We also measured the performance of node classification using F1-macro scores and observed that the relative performance of various methods is similar to the performance observed under F1-micro scores.

For some heterogeneous graph such as Youtube, r\toolname{} may perform better in the node classification task.  
For these graphs, multi-hop neighbors reached by random walks are able to capture complex relationships better. 
We tested r\toolname{} for Youtube with walk length set to 5.  
We observe that r\toolname{}, DeepWalk, HARP and Verse achieve an F1-micro score of 42\%, 45\%, 44\%, and 42\%, respectively, for 25\% training samples. DeepWalk performs better than other tools for this graph and r\toolname{} shows competitive performance. 


\subsection{Parameter sensitivity}
\label{sec:parametersense}
\begin{figure*}[!t]
    \centering
    \includegraphics[width=0.245\linewidth,height=3.6cm]{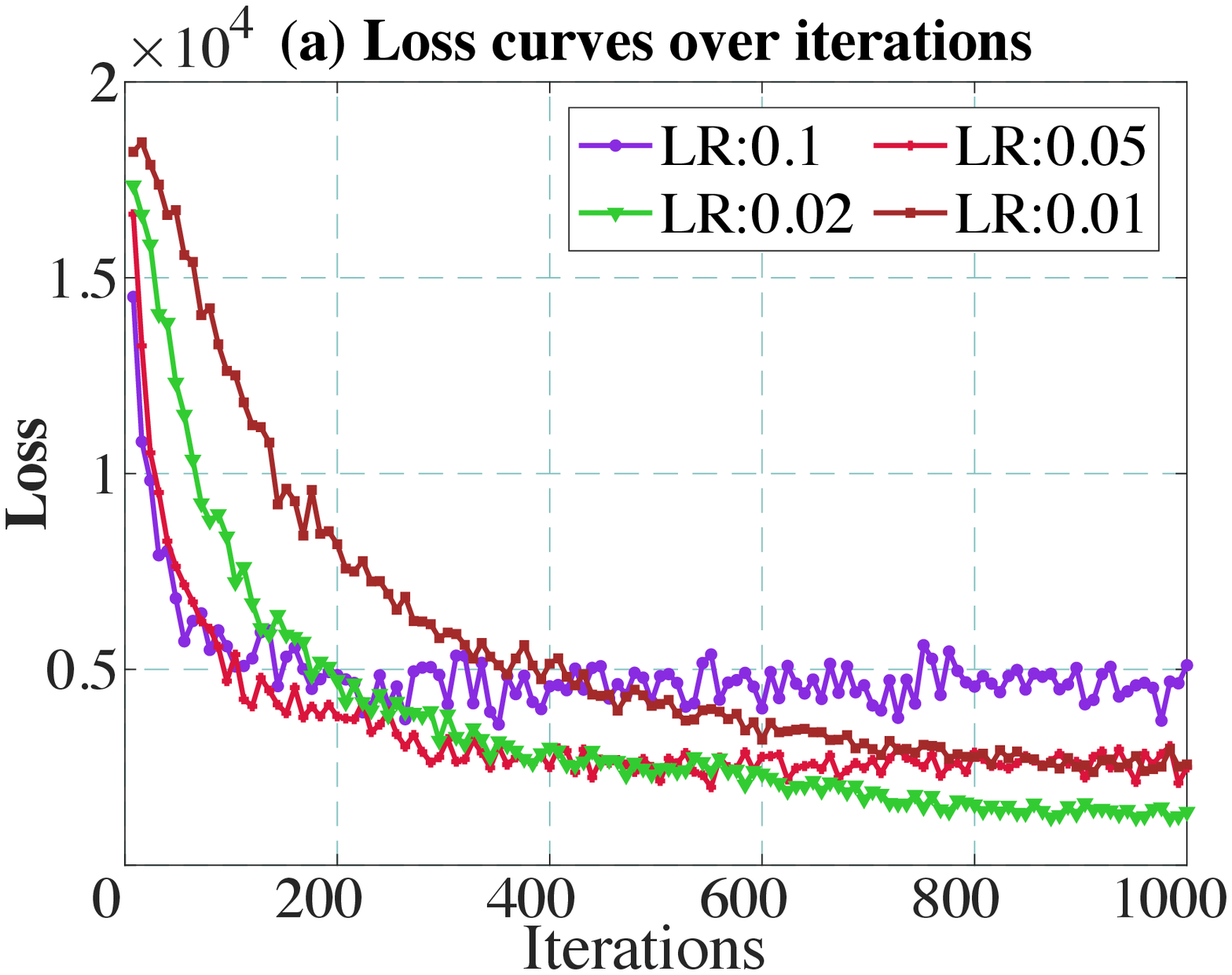}
    \includegraphics[width=0.245\linewidth,height=3.6cm]{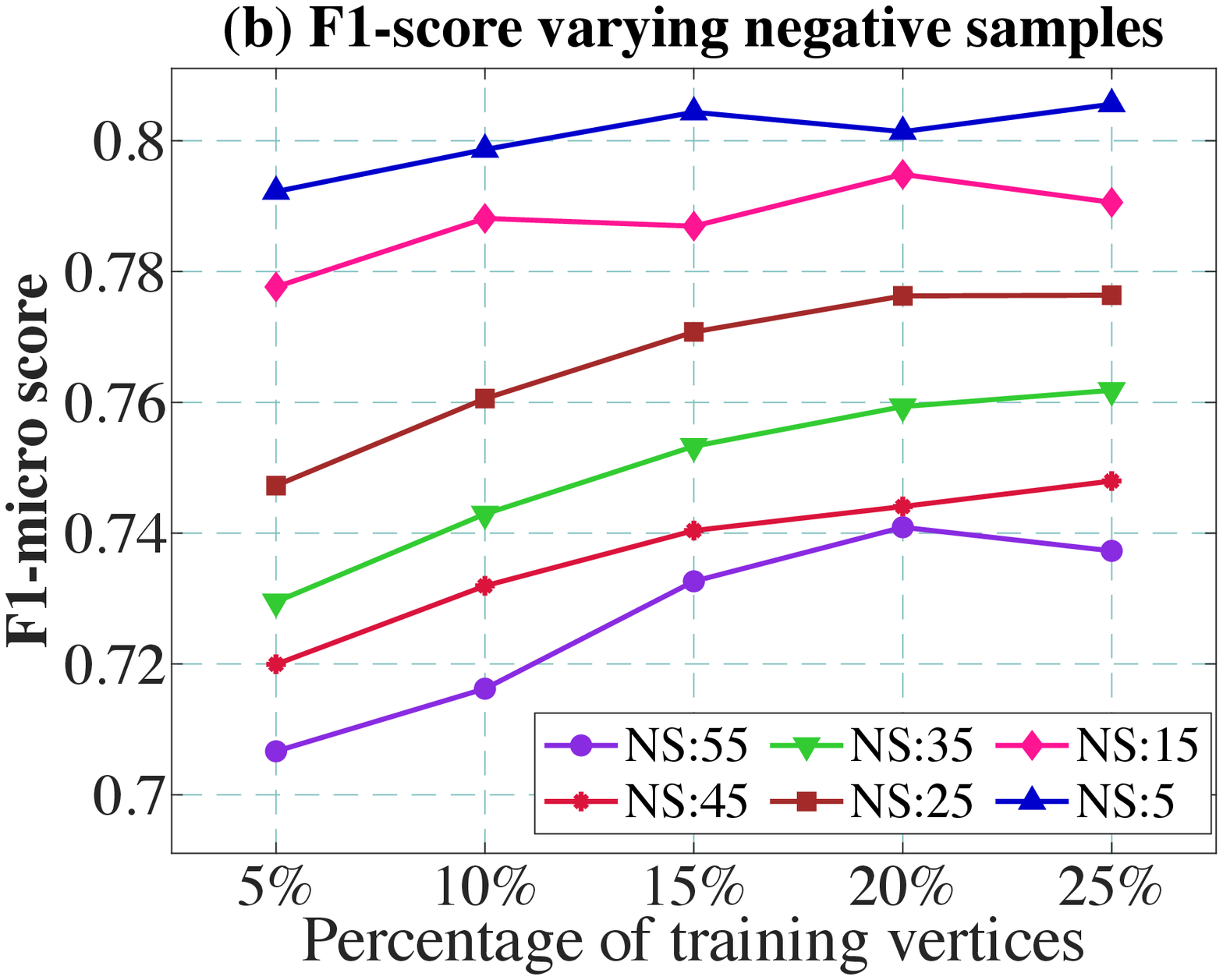}
    \includegraphics[width=0.245\linewidth,height=3.6cm]{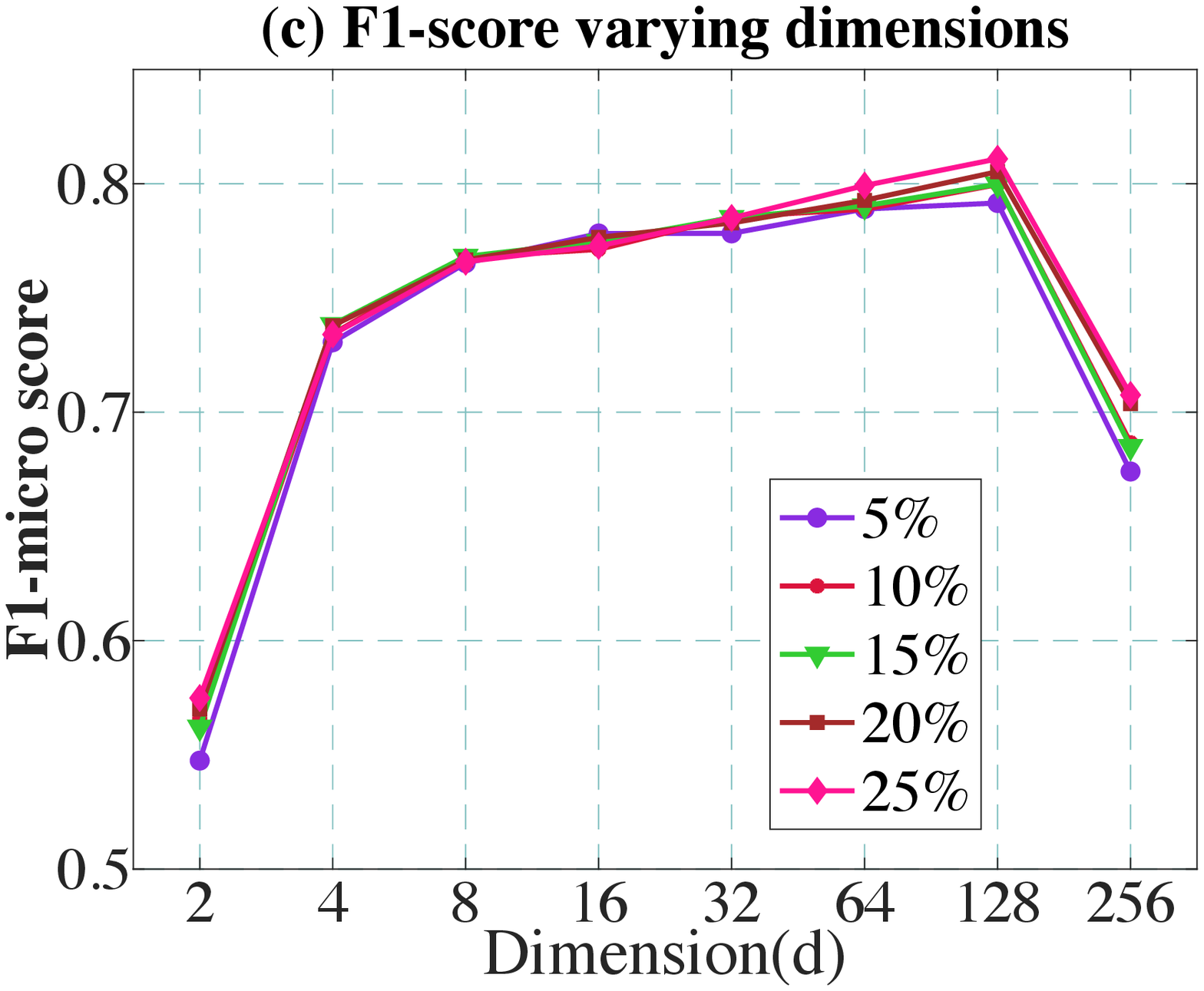}
    \caption{(a) Loss curves based on different learning rates for Pubmed dataset. (b) Effect of the number of negative samples on node classification task for Pubmed dataset. (c) Effect of dimensions for node classification task on Pubmed dataset. 
    }
    \vspace{-0.55cm}
    \label{fig:abcparametersensitivity}
\end{figure*}
We demonstrate parameter sensitivity of t\toolname{} using the Pubmed graph in Fig.~\ref{fig:abcparametersensitivity}. 

{\bf Convergence.} Fig.~\ref{fig:abcparametersensitivity}(a) shows loss curves of Eq. \ref{eqn:logloss} for different values of the learning rate $\eta$. 
A higher value of $\eta$ usually expedites the convergence, but the solution may get stuck to a local optima.  
By contrast, a small value of $\eta$ may lead to a better solution with a slow convergence rate. 
Hence, we keep $\eta$ as a hyper-parameters in \toolname{}. 
We empirically observed that $\eta = 0.02$ provides the best convergence-optima balance for  \toolname{} and thus, use it as a default value. 

{\bf Negative samples.} Fig.~\ref{fig:abcparametersensitivity}(b) shows F1-micro scores of node classification for different number of negative samples per vertex. We observe that \toolname{} shows better performance when we set $s$ to 5. 
This parameter is highly sensitive because a larger value of $s$ can 
create false negative samples that are also part of neighbors of vertices.

{\bf Embedding dimensions.}  Fig.~\ref{fig:abcparametersensitivity}(c) shows the results of node classification for different values of the embedding dimension $d$. 
As we increase $d$ from 2 to 256, the F1-micro score consistently increases up to $d=128$ and then starts decreasing.
Hence, we used $d=128$ for all previous experiments in the paper.
We also observe that F1-micro scores are remarkably stable for different percentages of training samples. 

{\bf Different force-directed models.} 
In addition to three variants of \toolname{} discussed thus far, we also implemented and experimented with other force-directed models mentioned in Table \ref{tab:diffmodels}, namely, Fruchterman Reingold (FR), ForceAtlas (FA), and LinLog (LL).
Since these models are directly used inside the gradient computation in Algorithm~\ref{algo:forceforbatch}, their runtimes are similar to t\toolname{}.
The performance of different force models in ML and visualization tasks varies from one graph to another. 
A detail study of their performance is beyond the scope of this paper. 

\vspace{-0.40cm}
\section{Conclusions}
This paper presents \toolname{}, a parallel graph embedding algorithm based on force-directed graph layout models.  
\toolname{} advances the graph embedding field in the following ways: (a) by incorporating proven visualization models into ML optimizations, \toolname{} provides high quality visualizations of graphs; (b) by using multiple levels of parallelism, \toolname{} runs at least an order of magnitude faster than previous graph embedding algorithms; (c) \toolname{} is uniformly good at node classification, link predictions and clustering compared to other embedding methods; (d) \toolname{} provides a generic and parallel framework that can be easily used with other force-directed and random-walk-based methods. 
Thus, \toolname{} enables large-scale graph mining and visualization in various scientific domains. 
\vspace{-0.20cm}
\section*{ACKNOWLEDGMENT}
 \RevisedText{We would like to thank anonymous reviewers for their feedback. Funding for this work was provided by the Indiana University Grand Challenge Precision Health Initiative.}

\bibliographystyle{IEEEtran}
\bibliography{references}


\end{document}